\documentclass[journal=jacsat,manuscript=article]{achemso}

\usepackage[version=3]{mhchem} %
\usepackage{soul}
\usepackage{color}
\usepackage{float}
\usepackage{graphicx}
\usepackage{caption}
\usepackage{subcaption}
\usepackage[section]{placeins}
\usepackage{array, makecell} %
\usepackage[colorlinks=true, citecolor=blue, linkcolor=blue, urlcolor=blue]{hyperref}

\author{Jianjun Hu}
\affiliation[USCUniversity]
{Department of Computer Science and Engineering, University of South Carolna, Columbia, SC, 29201, USA}
\email{jianunh@cse.sc.edu}

\author{Yong Zhao}
\affiliation[USCUniversity]{Department of Computer Science and Engineering, University of South Carolna, Columbia, SC, 29201, USA}
\altaffiliation{Current address: Visa Inc, Houston, Texas}

\author{Qin Li}
\affiliation[affiliationGZUFE]{
College of Big Data and Statistics,
  Guizhou University of Finance and Economics,
  Guiyang  550050,China
}
\author{Yuqi Song}
\affiliation[USCUniversity]{Department of Computer Science and Engineering, University of South Carolna, Columbia, SC, 29201, USA}
\author{Rongzhi Dong}
\affiliation[USCUniversity]{Department of Computer Science and Engineering, University of South Carolna, Columbia, SC, 29201, USA}

\author{ Wenhui Yang}

\affiliation[GZU]{School of Mechanical Engineering, Guizhou University, Guiyang,550050, China }

\author{Edirisuriya M. D. Siriwardane}
\affiliation[UColombo]{Department of Physics, University of Colombo, Colombo 3, Sri Lanka}

\title[An \textsf{achemso} demo]
  {Deep learning based prediction of contact maps and crystal structures of inorganic materials}

\abbreviations{IR,NMR,UV}

\begin{document}

\begin{tocentry}

TBD

\end{tocentry}

\begin{abstract}
Crystal structure prediction is one of the major unsolved problems in materials science. Traditionally, this problem is formulated as a global optimization problem for which global search algorithms are combined with first principle free energy calculations to predict the ground state crystal structure given only material composition. These ab initio algorithms are currently too slow for predicting complex material structures. Inspired by the AlphaFold algorithm for protein structure prediction, herein we propose AlphaCrystal, a crystal structure prediction algorithm that combines a deep residual neural network model for predicting the atomic contact map of a target material followed by 3D structure reconstruction using genetic algorithms. Extensive experiments on 20 benchmark structures showed that our AlphaCrystal algorithm can predict structures close to the ground truth structures, which can significantly speed up the crystal structure prediction and handle relatively large systems.

\end{abstract}

\section{Introduction}

The periodic crystal structures of inorganic materials determine the many unique and exotic functions of functional devices such as lithium batteries, quantum computers, solar panels, and chemical catalysts. While it is easy to compose a material with chemically reasonable formula or to generate millions of formulas with charge neutrality and electronegativity balance using modern generative machine learning algorithms such as MATGAN \cite{dan2020generative}, it is notoriously challenging to predict the crystal structure from a given chemical composition \cite{maddox1988crystals}, which however is required to check its thermodynamic and mechanical stability or their synthesizability \cite{jang2020structure,frey2019prediction, aykol2019network}. With the crystal structure of a chemical substance, many physichochemical properties can be predicted reliably and routinely using first-principle calculation or machine learning models \cite{xie2018crystal}.

Due to its importance in chemistry and condensed matter physics, crystal structure prediction has been investigated intensively for more than 30 years \cite{tsuneyuki1988first,bush1995evolutionary,glass2006uspex,woodley2008crystal,hautier2011data,wang2012calypso,curtis2018gator, avery2019xtalopt,ryan2018crystal,oganov2019structure}. 
In the crystal structure prediction (CSP) problem\cite{lyakhov2013new}, the goal is to find a ground state structure (in terms of all atomic coordinates of the atoms in a unit cell) with the lowest free energy for a given chemical composition (or a chemical system with variable composition) at a given pressure–temperature condition. It is assumed that atomic configurations with lower free energy correspond to more stable arrangement of atoms and the materials will be more synthesizable. One of the simplest and most widely used approaches for CSP is the template based or element substitution approach in which an existing crystal structure with a similar formula is first identified and then some atoms will be replaced with other type of elements. The replacement can either based on personal heuristics or guided by machine learned substitution rules in terms of element combination patterns\cite{fischer2006predicting,hautier2011data} or atomic fingerprints that describe coordination topology \cite{ryan2018crystal} or other chemical patterns \cite{shen2020charge} around unique crystallographic sites. Template based approaches are widely used in discovering new materials such as lithium ion cathode materials \cite{shen2020charge} and heteroanionic compounds \cite{he2020computational}. But they have a major limitation in its inability to generating new crystal structure types. 

The majority of work on the CSP problem is focused on ab initio approaches which try to search the atomic configuration space to locate the ground state structure guided by the first principle calculations of the free energy of candidate structures \cite{bush1995evolutionary,glass2006uspex,woodley2008crystal,hautier2011data,wang2012calypso,curtis2018gator, avery2019xtalopt}. These approaches use a variety of search/optimization algorithms such as random sampling,  simulated annealing, minima hopping, basin hopping, metadynamics, genetic algorithms, and partice swarm optimization to achieve systematic search while overcoming the local minima due to energy barriers in the search landscape. They have been successfully applied to discover a series of new materials as summarized in \cite{oganov2019structure, wang2020calypso}. To improve the sampling efficiency and save the costly DFT calculations, a variety of strategies have been proposed such as exploiting symmetry\cite{pretti2020symmetry} and pseudosymmetry\cite{lyakhov2013new}, smart variation operators, clustering, machine-learning interatomic potentials with active learning \cite{podryabinkin2019accelerating}. 

Despite their wide successes, the scalability and applicability of these ab initio CSP algorithms are severe limited due to their dependence on the costly DFT calculations of free energies for sampled structures. A quick check of their success stories reported in the literature \cite{oganov2019structure,zhang2017materials,wang2020calypso} can find that most of their discovered crystal materials are binary materials or those with less than 20 atoms in the unit cell. Our practice with these softwares shows that the algorithms tend to waste a lot DFT calculations to reach the local areas of the ground state structures, which may be addressed by seeding them with an approximate structure close to the target. With limited DFT calculations budget, how to efficiently sample the atom configurations becomes a key issue and the scalability of CSP remains an unsolved issue.

Here we propose a novel deep knowledge-guided ab initio approach for crystal structure prediction, which is inspired by the recent successes of deep learning approaches for protein structure prediction\cite{di2012deep,senior2020improved,zheng2019deep} led by the famous AlphaFold \cite{senior2020improved}. 
To our knowledge, our AlphaCrystal algorithm is the first method for crystal contact map prediction in CSP. We use deep residual neural networks \cite{he2016identity} for contact prediction which learns the intricate relationships of bonding relationships of atoms. The advantage of AlphaCrystal is that it can exploit the rich atom interaction distribution or other geometric patterns or motifs \cite{zhu2017efficient} existing in the large number of known crystal structures to predict the atomic contact map. These complex hidden knowledge can be learned as deep physical knowledge by our deep neural network, which can be exploited by the contact map prediction and atomic coordinate reconstruction process. We train the deep neural network using a subset of materials with solved structures from the Materials Project database and then test it on a set of test samples. Our experimental results show that our method when trained with 80\% MP samples can achieve almost 100\% contact map accuracy out of 48\% the test set.

Our contributions can be summarized as follows:

\begin{itemize}
  \item We propose AlphaCrystal, a deep learning and genetic algorithm based approach for crystal structure prediction using the predicted atomic contact map as a knowledge-guided methodology for addressing the crystal structure prediction problem.
  \item We evaluated our algorithm over 20 benchmark crystal targets and find that it can predict their contact maps with high accuracy, which further leads to successful prediction of their crystal structures.
  \item we have applied AlphaCrystal to predict 10 non-trivial crystal structures and verified their stability by DFT calculations.
\end{itemize}

\section{Methods}

\subsection{The AlphaCrystal framework for Contact map based crystal structure prediction}

\begin{figure*}[ht]
  \centering
  \includegraphics[width=0.85\linewidth]{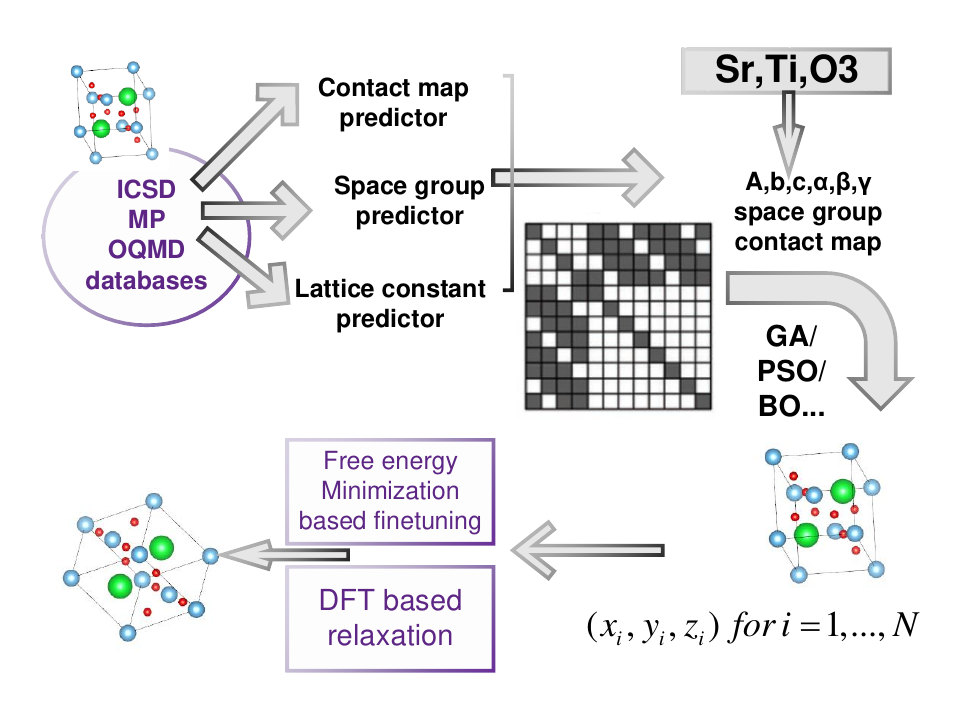}
  \caption{The AlphaCrystal framework for contact map based crystal structure prediction.}
  \label{fig:framework}
\end{figure*}

In our previous work \cite{hu2020contact}, we have shown that given a correct contact map along with the space group and lattice constants, a genetic algorithm can be used to reconstruct the atomic coordinates (so its structure) with high accuracy. Here we propose a deep learning based model for predicting the contact map given its composition only. Based on this contact map predictor, we propose AlphaCrystal, a new  framework for knowledge guided crystal structure prediction as shown in Figure\ref{fig:framework}. The architecture is composed of three main modules: 1) a contact map predictor, a space group predictor\cite{liang2020cryspnet}, and a lattice constant predictor\cite{li2020mlatticeabc}; 2) the contact map based atomic coordinate reconstruction algorithm\cite{hu2020contact}; 3) DFT relaxation based local search or free energy based ab initio search.  

\subsection{Deep learning model for crystal materials atomic contact prediction}

\begin{figure*}[ht]
  \centering
  \includegraphics[width=0.95\linewidth]{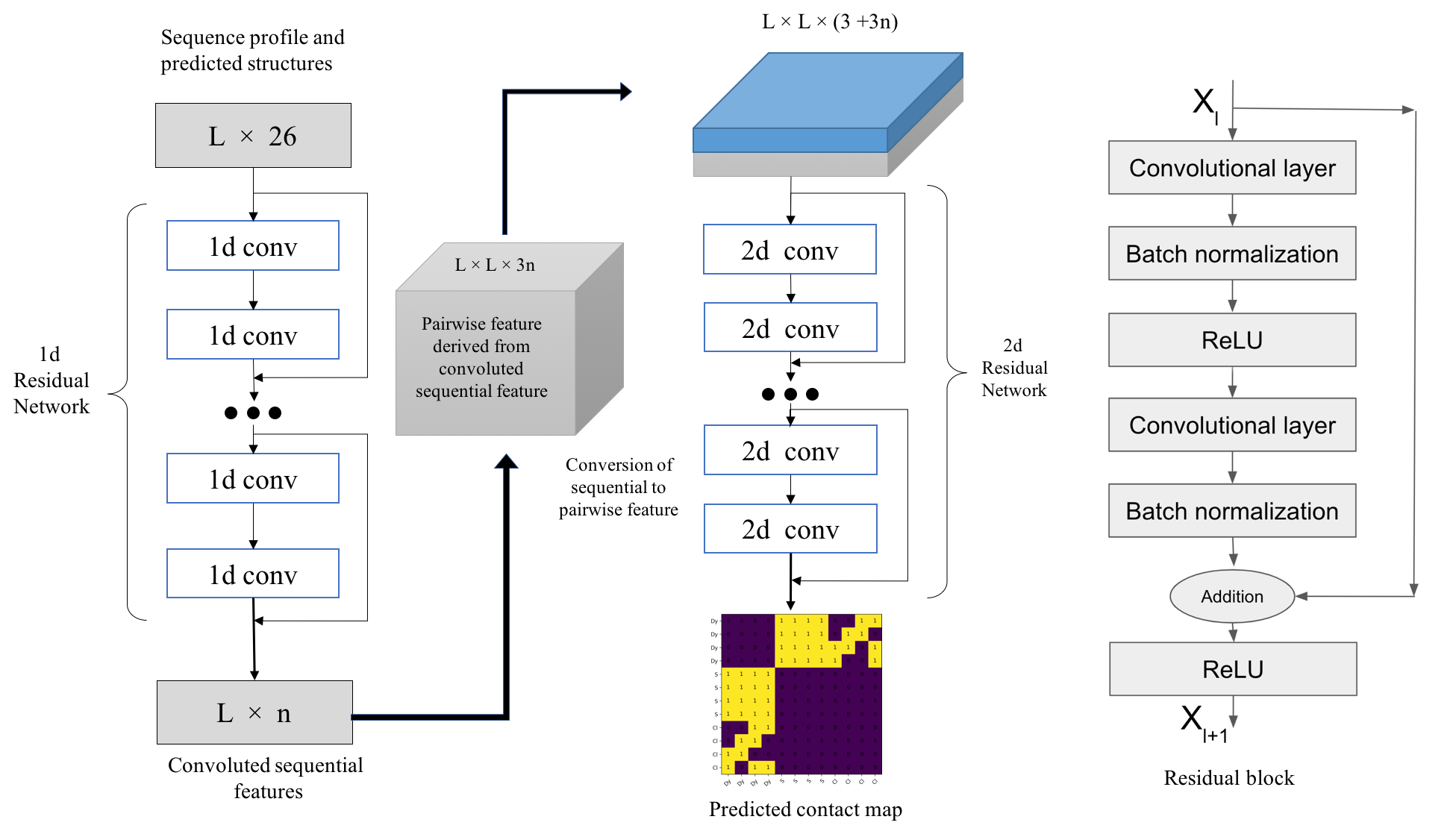}
  \caption{Deep neural network model for crystal material contact map prediction.}
  \label{fig:contactmappredictor}
\end{figure*}

One of the major components of AlphaCrystal algorithm is the deep residual network based predictor of contact maps. As shown in Figure\ref{fig:contactmappredictor}, the whole network is composed of three parts: the first part use a sequence of stacked 1D residual network layers to learn a convoluted atom site features. The input to this module is the sequence of element symbols in the input formula where L is the number of atoms in the unit cell. Each element is represented by 11 features including Mendeleev Number, unpaired electrons, ionization energies, covalent radius, heat of formation, dipole polarizability, average ionic radius, group number and row number in periodic table, pauling electronegativity, and atomic number. 

The second part of our contact map predictor is the conversion of convolved site features into pairwise feature maps with the dimension of $L\times L\times 3n$ (outer concatenation), where n indicates 128. The third module is composed of a sequence of stacked 2D residual network layers, which maps the paired sites features to predicted contact maps. Moreover, a batch normalization~\cite{ioffe2015batch} and a nonlinear transformation~\cite{krizhevsky2017imagenet} succeeds each convolutional layer. Batch normalization makes the training faster and stable by re-centering and re-scaling each layer at each mini-batch. The output of 1D residual network is an 2D matrix with dimension of $L\times n$, which is the learnt convoluted inter-atom site features hierarchically. The learnt inter-site features is converted to a 3D matrix as the inputs to 2D residual network.

\paragraph{Residual network block} Figure~\ref{fig:contactmappredictor} right pane shows the architecture of the residual network block used in our two residual network modules. In each block, there are two convolutional layers, a batch normalization and two nonlinear transformations. The nonlinear transformation is performed by the ReLU activation function $max(X, 0)$~\cite{krizhevsky2017imagenet}. Let $F(X_{l})$ denote the output of the block, and then $X_{l+1}$ is $max(F(X_{l}) + X_{l},0)$. The addition of $X_{l}$ and $F(X_{l})$ is non-linearly transformed. We use 9 building blocks for each module in our main architecture. The number of filters is doubled per 3 blocks. The initial number of filters for first and second module are 32 and 256, respectively.

\paragraph{Contact map generation and loss function}

We use the following rule to convert a crystal structure into a contact map matrix M: for each pair of atoms A,B in the unit cell, if their distance is within the range of $[CovalenceRadius_A + CovalenceRadius_B-0.4, CovalenceRadius_A + CovalenceRadius_B+0.4]$, then there is a bond between atom A and atom B and the corresponding M[i,j] is set to 1, otherwise it is set to 0. When two atoms are both metal atoms, we set their contact map entry as 0 too.

Since a contact map is a binary matrix, we use the cross-entropy loss as the loss function for neural network training. It is defined as: 

\begin{equation}
    Loss_{crossEntropy} =-\frac{1}{N} \sum_{i=0}^{N} y_{i} \cdot \log \left(\hat{y}_{i}\right)+\left(1-y_{i}\right) \cdot \log \left(1-\hat{y}_{i}\right)
\end{equation}
Where N is the maximum length of the formula which is set to 12 and 24 in our experiments; $y_{i}$ is the true contact map label at position i, and $\hat{y}_{i}$ is the predicted probability scores at position i.

\paragraph{Training and dealing with crystals of different atom sites}

To deal with varying sizes, we set the maximum number of atoms in a formula as L, which is set to 12 and 24 respectively in our experiments. When a formula has fewer atoms, we create the tensors by padding zeros. We sort all samples by their atom no, and then partition it into minibatch so that for each minibatch, the sizes are similar.

\subsection{Predictors for space group and lattice constants}

For each formula, we use CryspNet\cite{liang2020cryspnet} to predict top two crystal systems and top 5 space groups for each crystal system. We then use MLatticeABC \cite{li2020mlatticeabc} to predict the lattice constants for each formula. Next, we use the deep neural network model as shown in Figure \ref{fig:contactmappredictor} to predict the contact map.

\subsection{3D crystal structure reconstruction algorithm}

With all the predicted information including the contact map, the space group, and the lattice constants, we then used the CMCrystal \cite{hu2020contact}, a genetic algorithm for contact map based atomic position reconstruction, to predict the crystal structure for a given formula. We set the number of evaluations to be 100,000 or 1000 generations for a population size of 100 of the GA. The mutation rate is set to 0.001. Compared to previous version of the CMCrystal algorithm, we have added an additional term into the GA optimization objective function, which is the fitness of valid bonds. It is defined as:

\begin{equation}
    fitness_{bond}=\frac{No.of valid bonds}{No.of valid bonds+ No. of short bonds}
\end{equation}
where, short bonds are defined as any bond with a length less than the sum of two neighbor atoms' covalent radius minus 0.4{\AA}; valid bonds are those with length within the range of $[CovalenceRadius_A + CovalenceRadius_B-0.4, CovalenceRadius_A + CovalenceRadius_B+0.4]$ for atom pair A and B. The final fitness is defined as the product of contact map fitness and valid bond fitness.

\subsection{Evaluation metrics}

The objective function for contact map based structure reconstruction is defined as the dice coefficient, which is shown in the following equation:

\begin{equation}
\operatorname{fitness}_{opt}=\operatorname{Dice}=\frac{2|A \cap B|}{|A|+|B|} \approx\frac{2 \times A \bullet B}{\operatorname{Sum}(A)+\operatorname{Sum}(B)}
\end{equation}

where$A$ is the predicted contact map matrix and $B$ is the true contact map of a given composition, both only contain 1/0 entries.  $A \cap B$  denotes the common elements of A and B, |g| represents the number of elements in a matrix, • denotes dot product, Sum(g) is the sum of all matrix elements. Dice coefficient essentially measures the overlap of two matrix samples, with values ranging from 0 to 1 with 1 indicating perfect overlap. We also call this performance measure as contact map accuracy.

To evaluate the reconstruction performance of different algorithms, we can use the dice coefficient as one evaluation criterion, which however does not indicate the final structure similarity between the predicted structure and the true target structure. To address this, we define the root mean square distance (RMSD) and mean absolute error (MAE) of two structures as below:
\begin{equation}
    \begin{aligned}
\mathrm{RMSD}(\mathbf{v}, \mathbf{w}) &=\sqrt{\frac{1}{n} \sum_{i=1}^{n}\left\|v_{i}-w_{i}\right\|^{2}} \\
&=\sqrt{\frac{1}{n} \sum_{i=1}^{n}\left(\left(v_{i x}-w_{i x}\right)^{2}+\left(v_{i y}-w_{i y}\right)^{2}+\left(v_{i z}-w_{i z}\right)^{2}\right)}
\end{aligned}
\end{equation}

\begin{equation}
        \begin{aligned}
\mathrm{MAE}(\mathbf{v}, \mathbf{w}) &=\frac{1}{n} \sum_{i=1}^{n}\left\|v_{i}-w_{i}\right\| \\
&=\frac{1}{n} \sum_{i=1}^{n}\left(\|v_{i x}-w_{i x}\|+\|v_{i y}-w_{i y}\|+\|v_{i z}-w_{i z}\|\right)
\end{aligned}
\end{equation}

where $n$ is the number of independent atoms in the target crystal structure. For symmetrized cif structures, $n$ is the number of independent atoms of the set of Wyckoff equivalent positions. For regular cif structures, it is the total number of atoms in the compared structure. $v_i$ and $w_i$ are the corresponding atoms in the predicted crystal and the target crystal structure. It should be pointed out that in the experiments of this study, the only constraints for the optimization is the contact map, it is possible that the predicted atom coordinates are oriented differently from the target atoms in terms of of coordinate systems. To avoid this complexity, we compare the RMSD and MAE for all possible coordinate systems matching such as (x,y,z -->x,y,z), (x,y,z -->x,z,y), etc. and report the lowest RMSD and MAE. 

We also calculate a root mean square distances (RMS) as a performance measure of the structure prediction using the Pymatgen's structure matcher module. We use the $get_rms_dist$ function with fractional length tolerance $ltol=0.6$, site tolerance $stol=0.6$, and angle tolerance in degrees $angle_tol=20$ to compute the displacement between two structures. These threshold values are much larger than the defaults due to the range of discrepancy of the predicted structures and the ground truth ones.

\subsection{DFT validation of predicted structures}
 The predicted structures were relaxed using density functional theory (DFT) based on the  Vienna \textit{ab initio} simulation package (VASP)  \cite{Vasp1,Vasp2,Vasp3,Vasp4} in which projected augmented wave (PAW) pseudopotentials are implemented  \cite{PAW1, PAW2}. The plane-wave cutoff energy of 400 eV was considered with the Perdew-Burke-Ernzerhof (PBE) exchange-correlation functional of the generalized gradient approximation (GGA) \cite{GGA1, GGA2}. The structural optimization was performed energy and force criteria 1.0$\times$ 10$^{-5}$ eV/atom and 10$^{-2}$ eV/Å, respectively. The Brillouin zone integrations were carried out with $\Gamma$-centered  Monkhorst-Pack $k$-meshes.

\section{Results}
\label{sec:headings}

\subsection{Training and test data}

The contact map predictor in the present study is trained and tested using the MP database which is a database of inorganic crystal structures
with DFT-calculated properties consisting of almost all elements in
the periodic table and is freely accessible through the REST API interface. A total of 126,336 unique crystal structure data points queried in November 2020 (consisting of 46 781 synthesized crystals associated with the ICSD identifiers and 77734 theoretically proposed virtual crystals) were used for our learning model.

Training dataset is downloaded from Materials Project using Pymatgen API. We only choose crystal structure with the least formation energy if the corresponding formula has multiple structure. Materials with only metal elements are removed in this manuscript. We set the maximum number of atoms in the unit cell to be 12, which contains 11,355 samples.

\begin{figure}[tbh]
  \centering
  \includegraphics[width=0.99\linewidth]{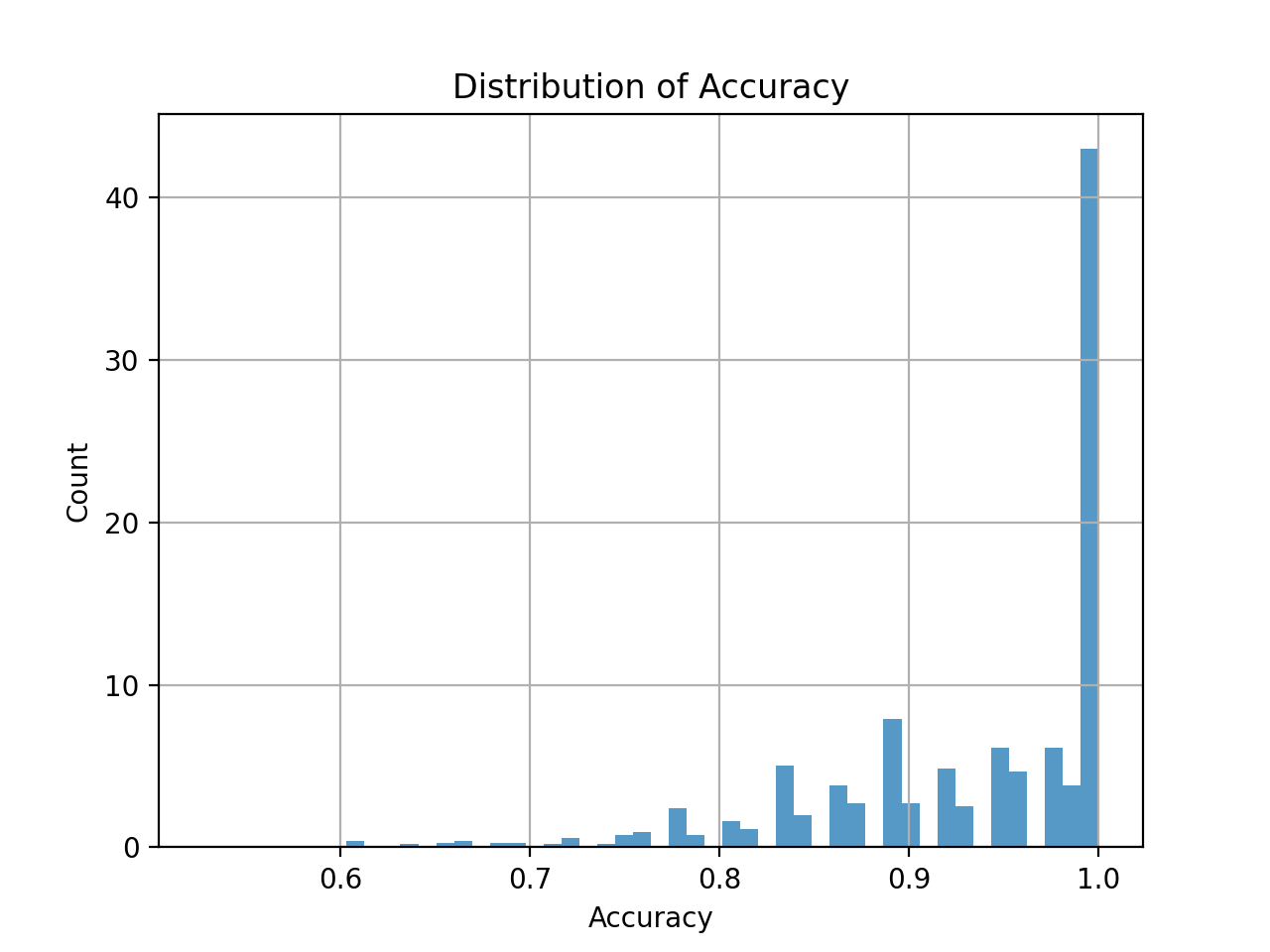}
  \caption{Contact map accuracy score distribution of predicted contact maps of test samples. For more than 40\% of the 1136 test samples, the contact map accuracy is 100\%.}
  \label{fig:dist-acc}
\end{figure}

\vspace{1cm}

\subsection{Overall contact map prediction performance}

\begin{figure*}[th] 
    \begin{subfigure}[t]{0.5\textwidth}
        \includegraphics[width=\textwidth]{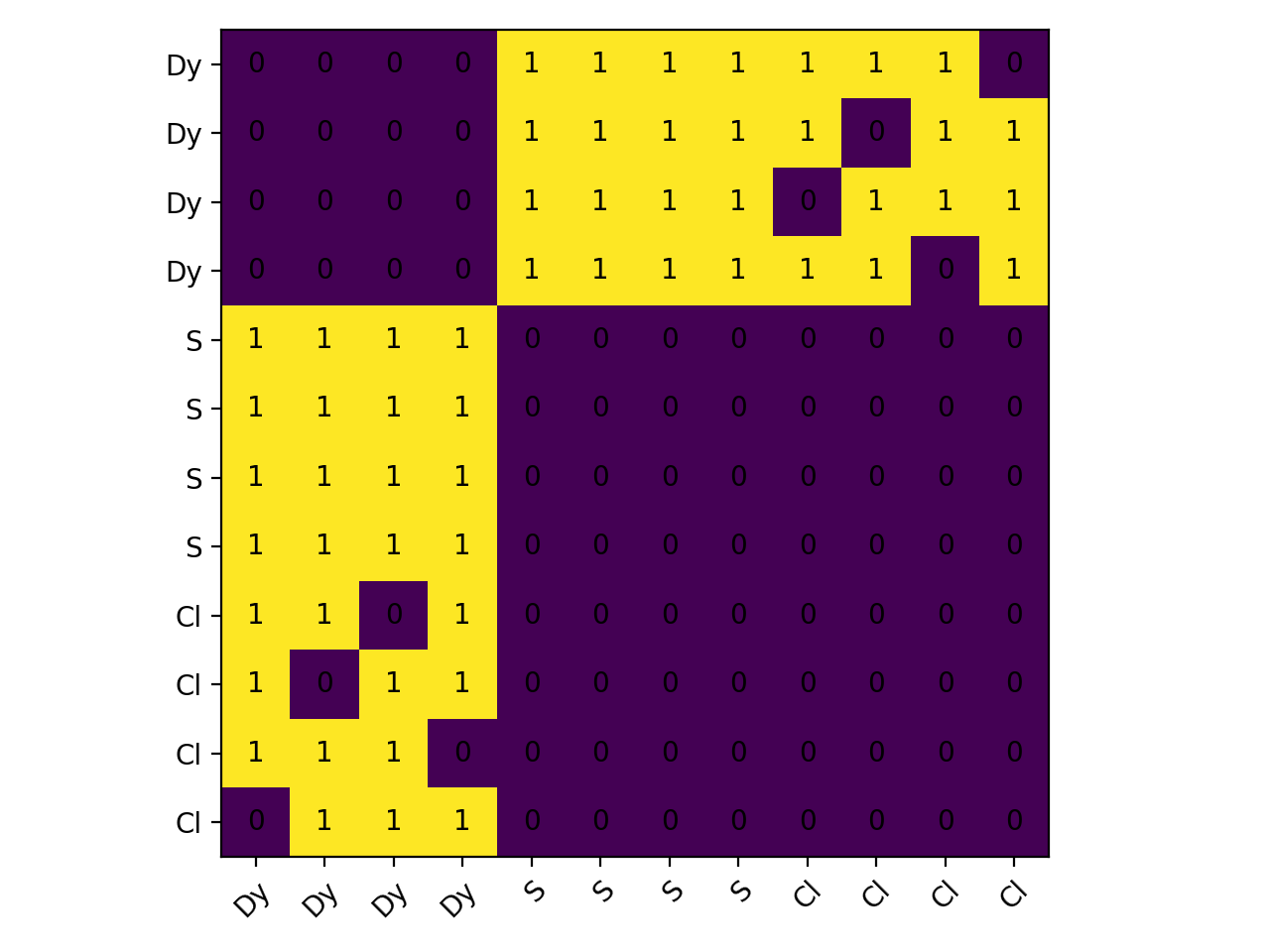}
        \caption{Dy4S4Cl4 (real)}
        \vspace{-3pt}
        \label{fig:t-Dy4S4Cl4}
    \end{subfigure}\hfill
    \begin{subfigure}[t]{0.5\textwidth}
        \includegraphics[width=\textwidth]{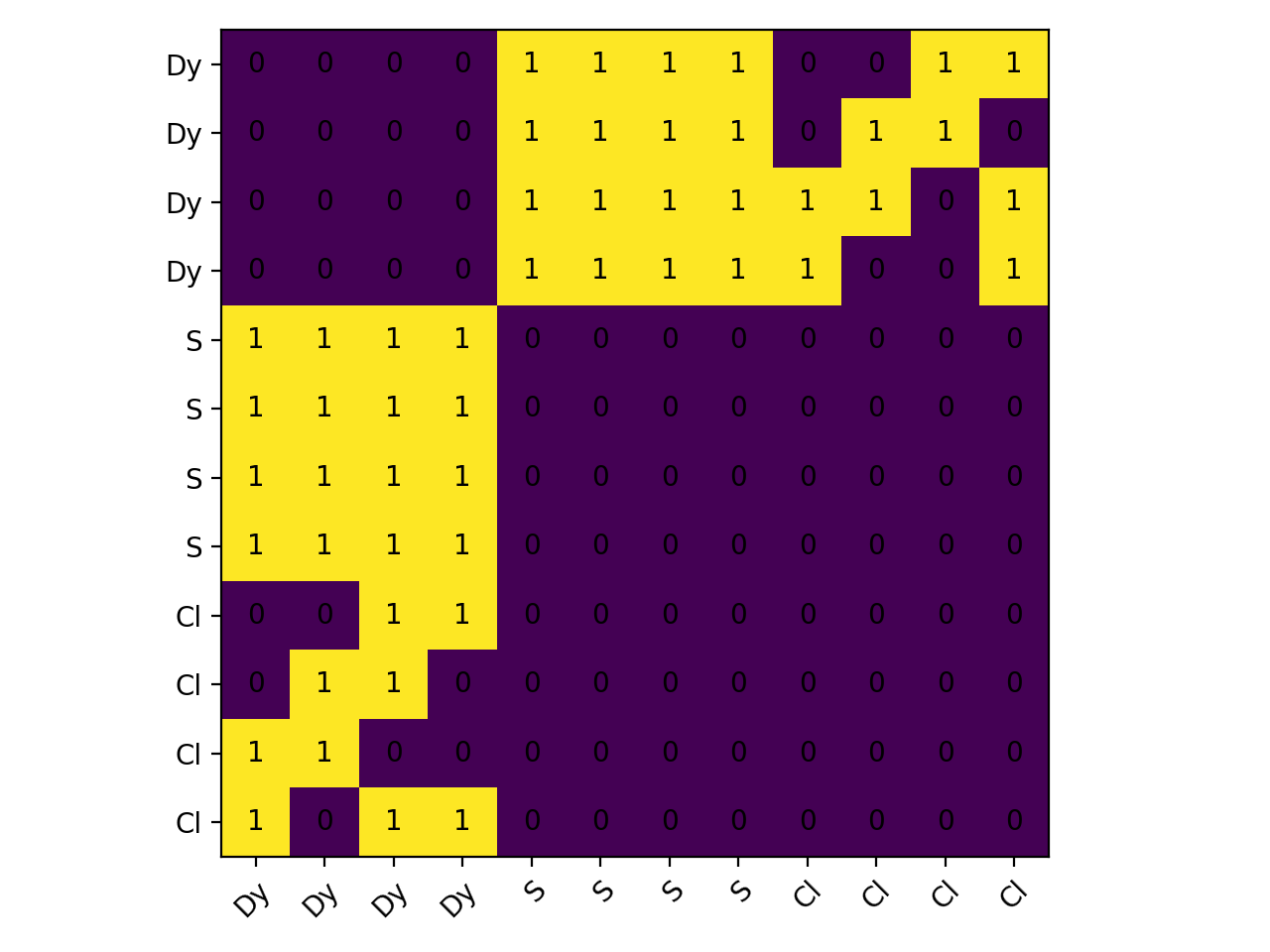}
        \caption{Dy4S4Cl4 (predicted)}
        \vspace{-3pt}
        \label{fig:p-Dy4S4Cl4}
    \end{subfigure}
    
    \begin{subfigure}[t]{0.5\textwidth}
        \includegraphics[width=\textwidth]{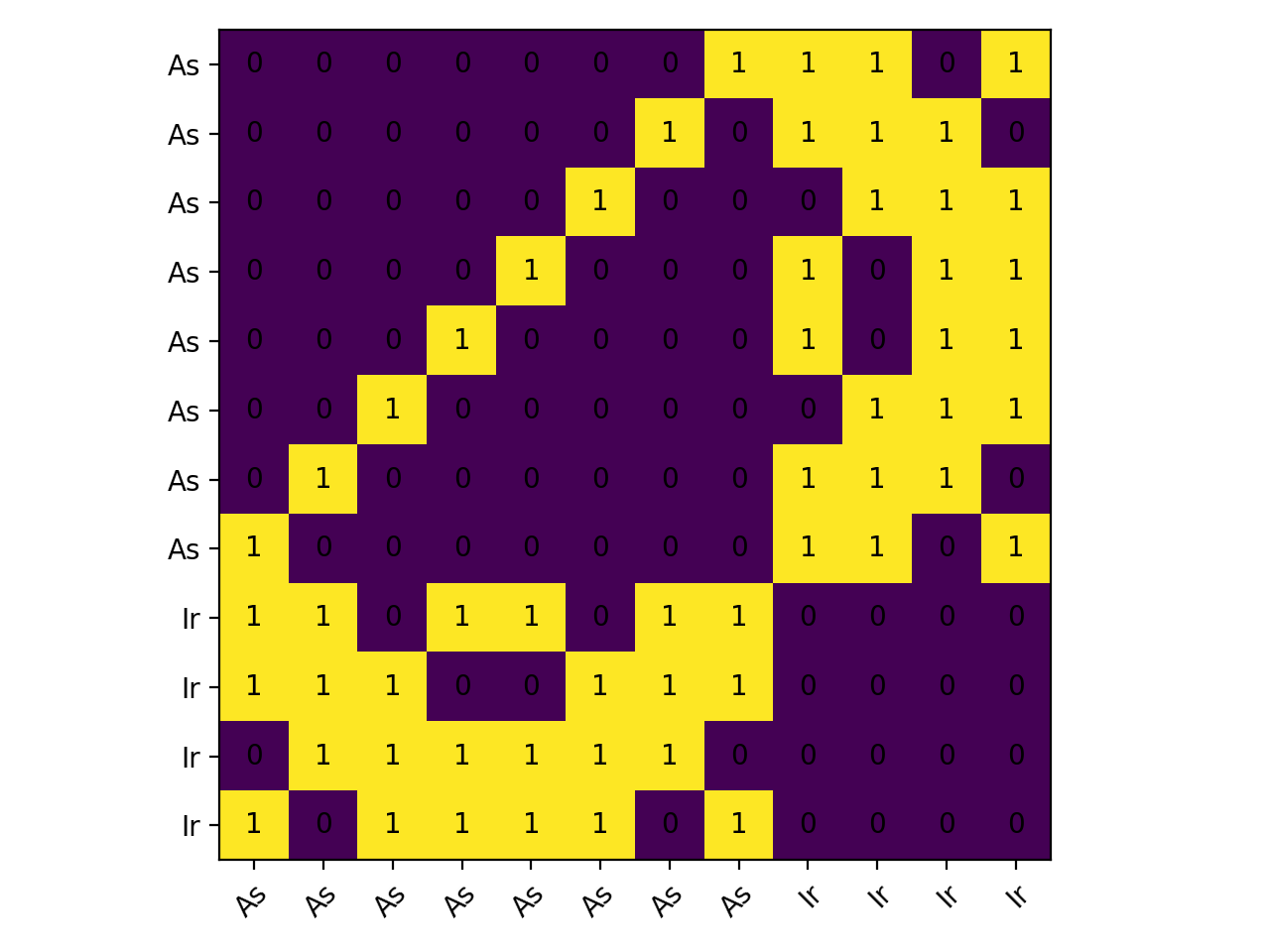}
        \caption{As8Ir4 (real)}
        \vspace{-3pt}
        \label{fig:t-As8Ir4}
    \end{subfigure}\hfill
    \begin{subfigure}[t]{0.5\textwidth}
        \includegraphics[width=\textwidth]{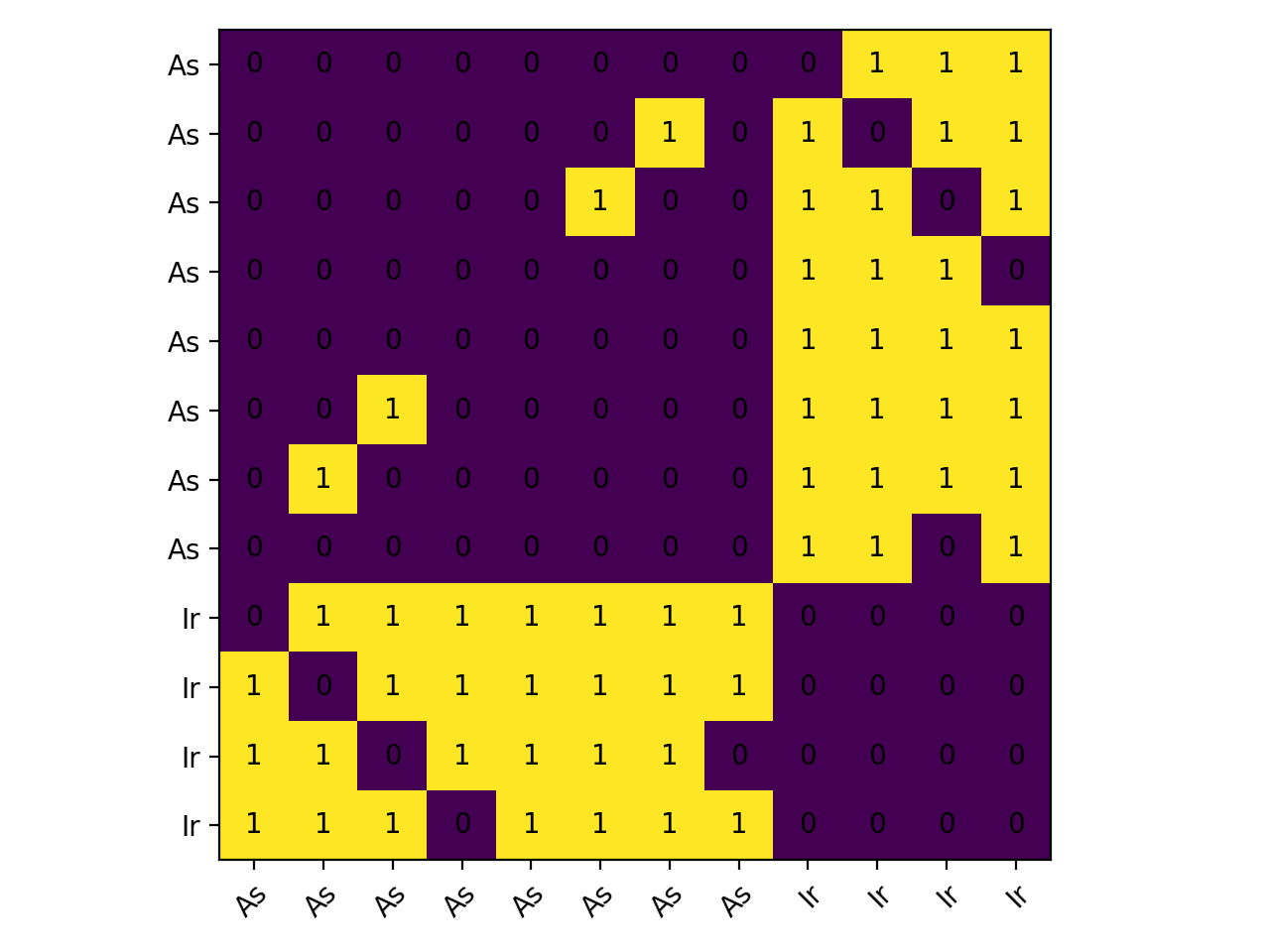}
        \caption{As8Ir4 (predicted)}
        \vspace{-3pt}
        \label{fig:p-As8Ir4}
    \end{subfigure}
    \caption{Contact maps for Dy4S4Cl4 and As8Ir4. Yellow cells indicate bonds while black cells show no bonds.}
    \label{fig:contact_maps}
\end{figure*}

In this experiment, we hold out 1136 known materials in the dataset as the test set and use the remaining 10219 samples as the training set for training our deep neural network model for contact map prediction. We set the number of epochs to be 125, Adam optimizer~\cite{kingma2014adam} is used to update model parameters and the learning rate to be 0.0001. After training, the model is used to predict the contact maps of the test samples and their contact map accuracy scores are plotted in Figure\ref{fig:dist-acc}. The average and standard deviation of prediction accuracy are 0.927 and 0.090, respectively. It is impressive to see that almost 461 out of 1136 test samples have then contact map predicted with 100\% accuracy. For more than 91\% of test samples, the contact map accuracy is higher than 80\%, indicating that the deep contact map predictor has captured the bonding relationships of atoms in the crystal structures. Figure\ref{fig:contact_maps} shows two examples of the true contact maps and the predicted contact maps for Dy4S4Cl4 and As8Ir4.

\vspace{1cm}

To further examine the contact map prediction performance, Table \ref{table:overall_performance} shows the contact map accuracy for 10  structures of different space groups with different number of atoms ranging from 6 to 12 atoms in their unit cells. We find for binary materials, the contact map accuracy scores range from 0.6 to 1.0 for Mg\textsubscript{2}P\textsubscript{8} and Ag\textsubscript{2}F\textsubscript{4} respectively. The number of independent sites are not the only determining factor as Pd\textsubscript{4}S\textsubscript{8} has only two independent sites but its contact accuracy is 0.694 which is lower than the other binary structures with three independent sites such as As\textsubscript{8}Ir\textsubscript{4} and Ge\textsubscript{4}F\textsubscript{8}. For ternary materials, Table \ref{table:overall_performance} shows that our model can also achieve high contact map accuracy over Si\textsubscript{4}Pt\textsubscript{4}Se\textsubscript{4}  and Ta\textsubscript{4}N\textsubscript{4}O\textsubscript{4}. 

\vspace{1cm}
\FloatBarrier
\begin{table*}[ht] 
\begin{center}
\caption{Performances of AlphaCrystal in terms of contact map prediction accuracy}
\label{table:overall_performance}
\begin{tabular}{|l|l|l|l|l|l|l|}
\hline
\multicolumn{1}{|c|}{\textbf{Target}} & \multicolumn{1}{c|}{\textbf{mp\_id}} & \textbf{\makecell{No.of \\sites}} & \multicolumn{1}{c|}{\textbf{\makecell{atom\# in\\unit cell}}} & \textbf{\makecell{\# of\\ variables}} & \multicolumn{1}{c|}{\textbf{\makecell{space\\ group}}} & \multicolumn{1}{c|}{\textbf{\makecell{contact map \\accuracy}}}  \\ \hline
Ag\textsubscript{2}F\textsubscript{4}             & 7715           & 2     & 6               & 6        & 14       & 1.0                       \\ \hline
Mg\textsubscript{2}P\textsubscript{8}             & 384          & 3    & 10               & 9        & 14       & 0.6                    \\ \hline
Ru\textsubscript{2}F\textsubscript{8}              & 974434           & 3    & 10               & 9        & 14      & 0.86                        \\ \hline
As\textsubscript{8}Ir\textsubscript{4}              & 15649            & 3    & 12               & 9        & 14      & 0.819                     \\ \hline
Ge\textsubscript{4}F\textsubscript{8}            & 7595            & 3    & 12               & 9        & 19      & 0.819                       \\ \hline
Pd\textsubscript{4}S\textsubscript{8}             & 13682           & 2    & 12               & 6        & 61     & 0.694                       \\ \hline
Dy\textsubscript{4}S\textsubscript{4}Cl\textsubscript{4}              & 561307           & 3    & 12               & 9        & 14     & 0.875                       \\ \hline
Si\textsubscript{4}Pt\textsubscript{4}Se\textsubscript{4}            & 1103261           & 3    & 12               & 9        & 29     & 1.0                      \\ \hline
Ta\textsubscript{4}N\textsubscript{4}O\textsubscript{4}            & 4165            & 3    & 12               & 9        & 14     & 1.0                        \\ \hline
\end{tabular}
\end{center}
\end{table*}

\subsection{Contact map based crystal structure prediction: benchmark results}

\begin{figure*}[htb!]
	\centering

	\begin{subfigure}{.3\textwidth}
		\includegraphics[width=\textwidth]{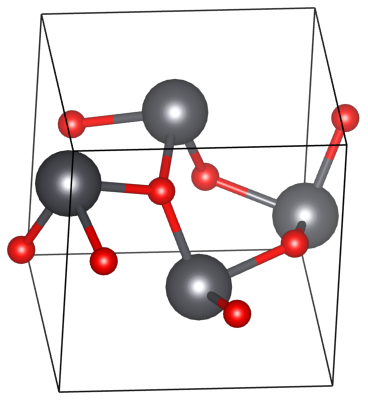}
		\caption{Target structure Pb\textsubscript{4}O\textsubscript{4}}
		\vspace{3pt}
	\end{subfigure}
		\begin{subfigure}{.33\textwidth}
		\includegraphics[width=\textwidth]{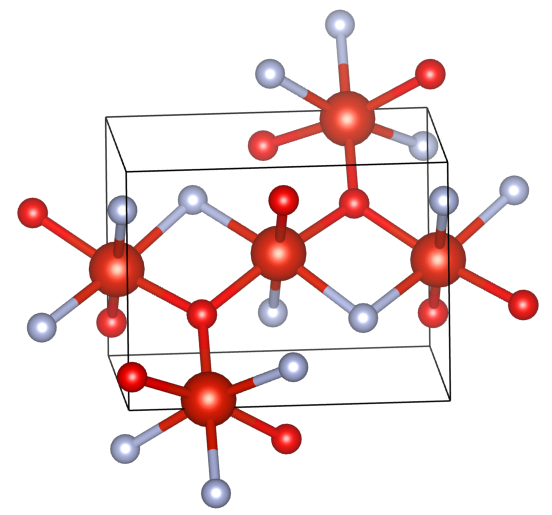}
		\caption{Target structure V\textsubscript{4}O\textsubscript{4}F\textsubscript{4}}
		\vspace{3pt}
	\end{subfigure}
		\begin{subfigure}{.3\textwidth}
		\includegraphics[width=\textwidth]{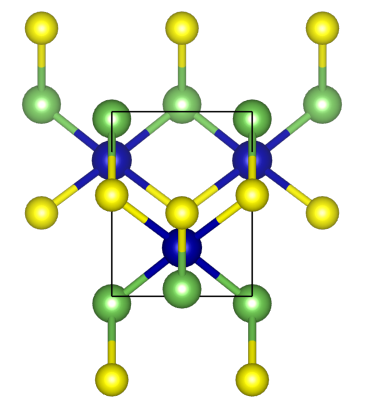}
		\caption{Target structure Co\textsubscript{2}As\textsubscript{2}S\textsubscript{2}}
		\vspace{3pt}
	\end{subfigure}
	
		\begin{subfigure}{.28\textwidth}
		\includegraphics[width=\textwidth]{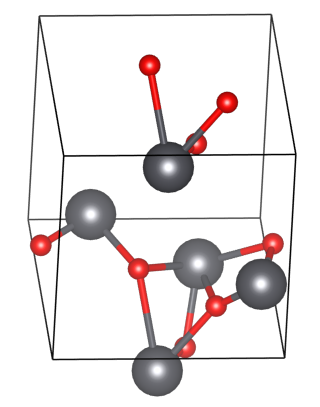}
		\caption{Predicted structure Pb\textsubscript{4}O\textsubscript{4}}
		\vspace{3pt}
	\end{subfigure}
		\begin{subfigure}{.28\textwidth}
		\includegraphics[width=\textwidth]{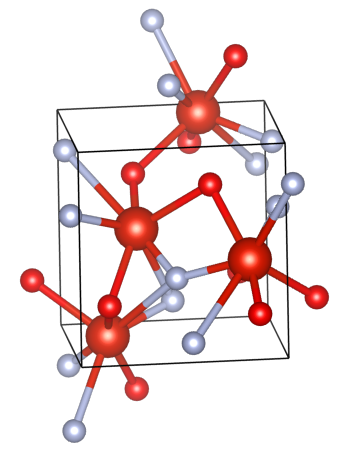}
		\caption{Predicted structure V\textsubscript{4}O\textsubscript{4}F\textsubscript{4}}
		\vspace{3pt}
	\end{subfigure}
		\begin{subfigure}{.35\textwidth}
		\includegraphics[width=\textwidth]{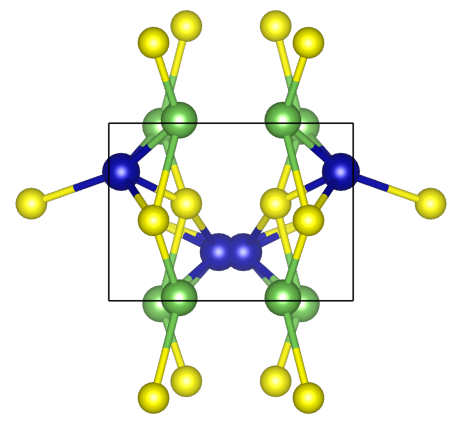}
		\caption{Predicted structure Co\textsubscript{2}As\textsubscript{2}S\textsubscript{2}}
		\vspace{3pt}
	\end{subfigure}
	\caption{Benchmark crystal structures predicted by AlphaCrystal.(a-c) target structures (d) predicted structure of Pb\textsubscript{4}O\textsubscript{4} with contact map accuracy:95.24\%, RMSD:0.376; (e) predicted structure of V\textsubscript{4}O\textsubscript{4}F\textsubscript{4} with contact map accuracy:88.89\%, RMSD:0.196;(f) Predicted structure of Co\textsubscript{2}As\textsubscript{2}S\textsubscript{2} with contact map accuracy:85.71\%, RMSD:0.196 }
	\label{fig:benchmarkstructures}
\end{figure*}

Here we evaluate how the predicted contact maps by our deep neural network model can be used to edit the crystal structures using our CMCrystal algorithm \cite{hu2020contact}, in which a genetic algorithm is used to search the fractional coordinates of the crystals with specified space group and contact map by minimizing the contact map distance error. We select 10 target structures as shown in Table \ref{table:structure_performance}, and then predict their contact maps using our neural network model. Next, we use the CryspNet algorithm to predict their crystal systems and top 5 space groups. We then use the MLatticeABC to predict the lattice parameters. However, we find that the CryspNet is not reliable enough to always predict the ground truth space group as its top 5 predictions. Considering that there are only a limited number of space groups, it is possible to exhaustively use each of the possible space group combined with the predicted contact map and lattice parameters to reconstruct the structures and then pick the structure with the lowest DFT-calculated formation energy. This can be done using 270 jobs (corresponding to 270 possible space groups) on Linux clusters in parallel. For simplicity, here we just directly specify the ground truth space groups and combine them with the predicted contact maps and lattice parameters to do structure reconstruction using CMCrystal algorithm. Then we calculate the corresponding RMSD, MAE, and RMS errors for all the reconstructed structures as shown in Table\ref{table:structure_performance}(last three columns).

\begin{table*}[ht!] 
\small
\begin{center}
\caption{Structure prediction performance of AlphaCrystal with ground truth space groups}
\label{table:structure_performance}
\begin{tabular}{|l|l|l|l|l|l|l|l|l|l|}
\hline
\multicolumn{1}{|c|}{\textbf{Target}} & \multicolumn{1}{c|}{\textbf{mp\_id}} & \textbf{\makecell{atom\# \\in\\ unit\\cell}} & \multicolumn{1}{c|}{\textbf{\makecell{given \\space\\group}}} & \textbf{\makecell{target\\space\\group}} & \multicolumn{1}{c|}{\textbf{\makecell{predicted\\ contact\\ map\\ accuracy}}} & \multicolumn{1}{c|}{\textbf{\makecell{reconstruct \\contact\\ map \\accuracy}}} & \multicolumn{1}{c|}{\textbf{RMSD}} & \textbf{MAE} &\textbf{RMS}\\ \hline
Cr\textsubscript{3}O\textsubscript{5}             & 1096920           & 8     & 1               & 1        &  0.939      & 0.917            & 0.411             & 0.314     & 0.387        \\ \hline
Pb\textsubscript{4}O\textsubscript{4}              & 550714           & 8    & 29               & 29        &   0.696    & 0.952            & 0.376             & 0.324  & 0.398           \\ \hline
Co\textsubscript{4}P\textsubscript{8}             & 14285           & 12    & 14               & 14        &  0.727    & 0.968            & 0.196             & 0.156    & None         \\ \hline
Ir\textsubscript{4}N\textsubscript{8}            & 415           & 12    & 14               & 14        & 0.773     & 0.952            & 0.145     & 0.128        & None             \\ \hline
V\textsubscript{2}Cl\textsubscript{10}            & 1101909            & 12    & 2               & 2        &  0.848    & 0.889            & 0.212             & 0.156  & 0.466           \\ \hline
Co\textsubscript{2}As\textsubscript{2}S\textsubscript{2}             & 553946           & 6    & 31               & 31        & 0.939     & 0.857            & 0.196             & 0.146    & 0.404         \\ \hline
V\textsubscript{2}O\textsubscript{1}F\textsubscript{7}             & 765500           & 10     & 1               & 1        &  0.939      & 1.0            & 0.336             & 0.255     & None        \\ \hline
V\textsubscript{4}O\textsubscript{4}F\textsubscript{4}             & 754589          & 12    & 92               & 92        &   0.803     & 0.889            & 0.196             & 0.171   & 0.382          \\ \hline
Fe\textsubscript{4}As\textsubscript{4}Se\textsubscript{4}              & 1101894            & 12    & 14               & 14        &   1.0    & 1.0            & 0.193             & 0.163   & 0.531          \\ \hline
Mn\textsubscript{4}Cu\textsubscript{4}P\textsubscript{4}              & 20203           & 12    & 62               & 62        &  0.879    & 0.941            & 0.146    & 0.117         & 0.560             \\ \hline
\end{tabular}
\end{center}
\end{table*}

There are several interesting observations. First, we found that for all these binary and ternary benchmark materials, our algorithm has achieved good contact map prediction accuracy as shown in column 6 of Table\ref{table:structure_performance} which ranges from 0.696 to 1.0. The structure prediction performances are shown in column 8 with the lowest RMSD of 0.145 for Ir\textsubscript{r}N\textsubscript{8}. In term of MAE error, the best performance is on Mn\textsubscript{4}Cu\textsubscript{4}P\textsubscript{4} with a MAE error of 0.117 despite the predicted contact map accuracy is not the highest with a score of 0.879. We also tried to calculate the root mean square error as defined by the Pymatgen routine and find for some of the structures it cannot calculate successfully while for others, the distances are not consistent with our RMSE/MAE results possibly due to the deviations of the predicted structures are too large to the ground truth structures to calculate them using their algorithm. Another interesting observation is from comparison of the predicted contact map accuracy and the reconstruction contact map accuracy, which evaluates how the contact map from predicted structures matches the contact map from ground truth structures. We find that in most cases, the GA based optimization algorithms CMCrystal can achieve high accuracy to reconstruct the predicted contact maps with accuracy ranging from 0.857 to 1.0.

Figure \ref{fig:benchmarkstructures} shows three target structures and their predicted ones. For Pb\textsubscript{4}O\textsubscript{4}, our algorithm achieves a contact map accuracy of 95.24\% and RMSD of 0.376. For target structure V\textsubscript{4}O\textsubscript{4}F\textsubscript{4}, even though the contact map accuracy is lower (88.89\%), the RMSD error is lower with a score of 0.196 and the overall structures are similar. Similar RMSD score has also been achieved for target Co\textsubscript{2}As\textsubscript{2}S\textsubscript{2}.

\subsection{Discovery of new structures using AlphaCrystal}

In our previous work, we developed MatGAN\cite{dan2020generative}, a deep generative machine learning for large-scale generation of new hypothetical materials compositions with chemical validity and high potential of being stable. Here we use MATGAN to generate 5 million hypothetical materials compositions and then apply charge neutrality check and electronegativity balance check. Then we train a composition based formation energy predictor using Roost, a composition and graph based predictor. We then filter out those candidates with La or Ac elements. We use the trained free energy predictor to screen the top 100 compositions with the lowest predicted formation energies and with the number of atoms less than 12 and the number of elements in the compounds to be 2 or 3.

For the selected 100 candidate materials, we use the CryspNet\cite{liang2020cryspnet}, a composition and deep neural network predictor for crystal systems and space groups to predict the top 2 crystal systems. For each predicted crystal system, we predict the top 5 space groups for the candidate. So for each composition, we have 10 candidate structures of different space groups. For each of such structure candidates, we apply the MLatticeABC algorithm to predict its lattice constants $a,b,c$. 

\begin{figure*}[htb!]
	\centering

	\begin{subfigure}{.33\textwidth}
		\includegraphics[width=\textwidth]{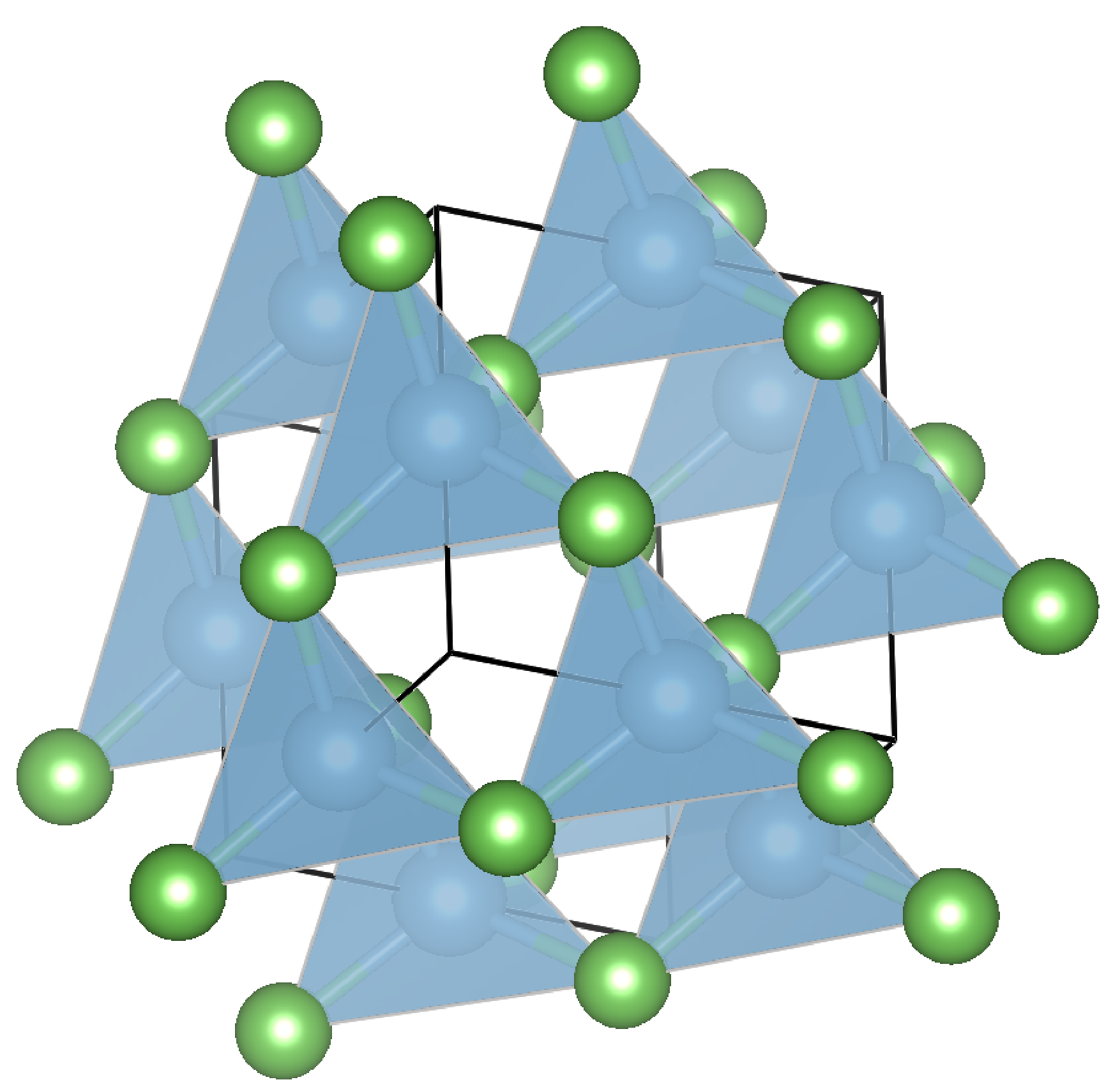}
		\caption{ }
		\vspace{3pt}
	\end{subfigure}
	\begin{subfigure}{.3\textwidth}
		\includegraphics[width=\textwidth]{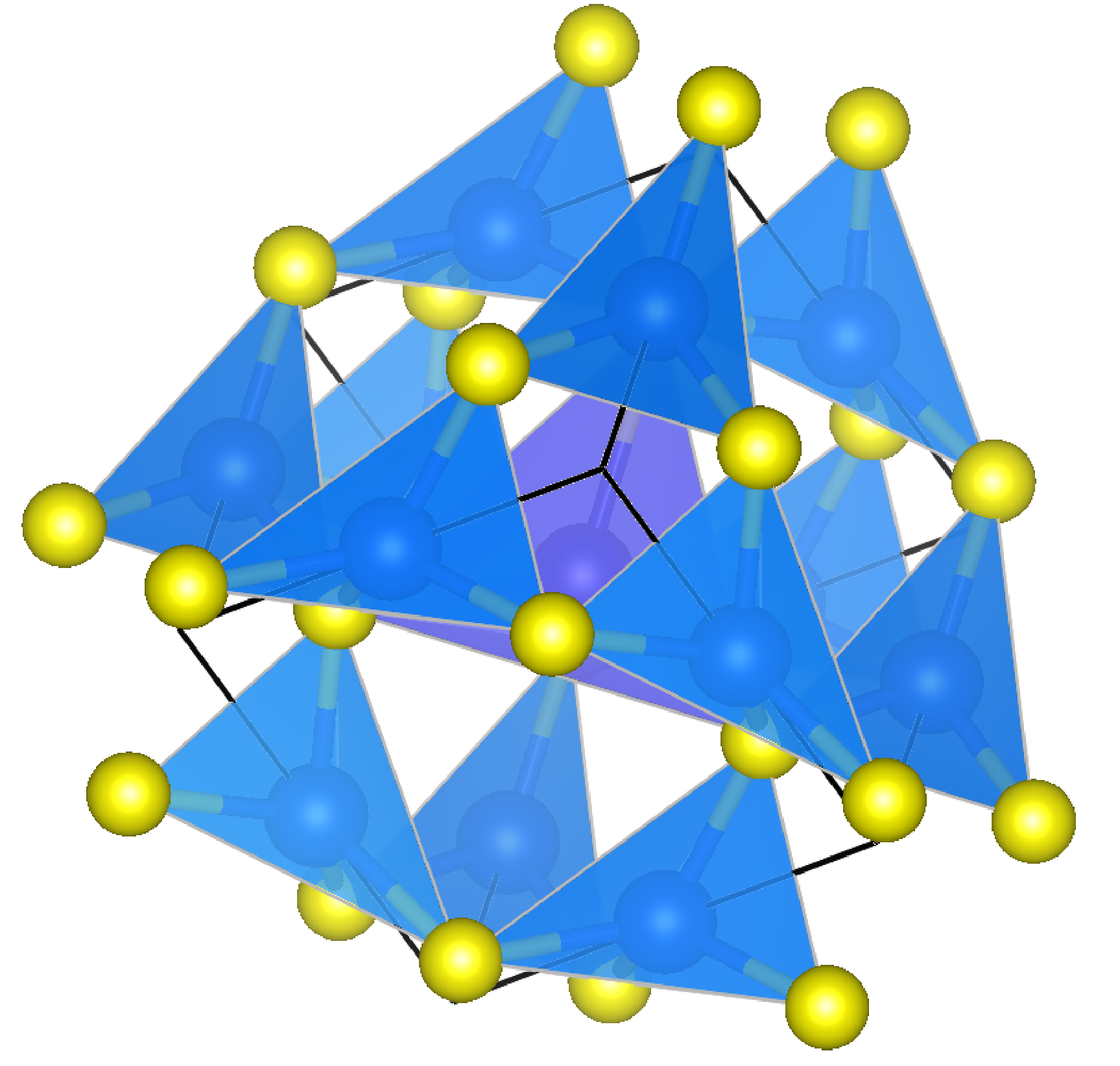}
		\caption{}
		\vspace{3pt}
	\end{subfigure}
	\begin{subfigure}{.33\textwidth}
		\includegraphics[width=\textwidth]{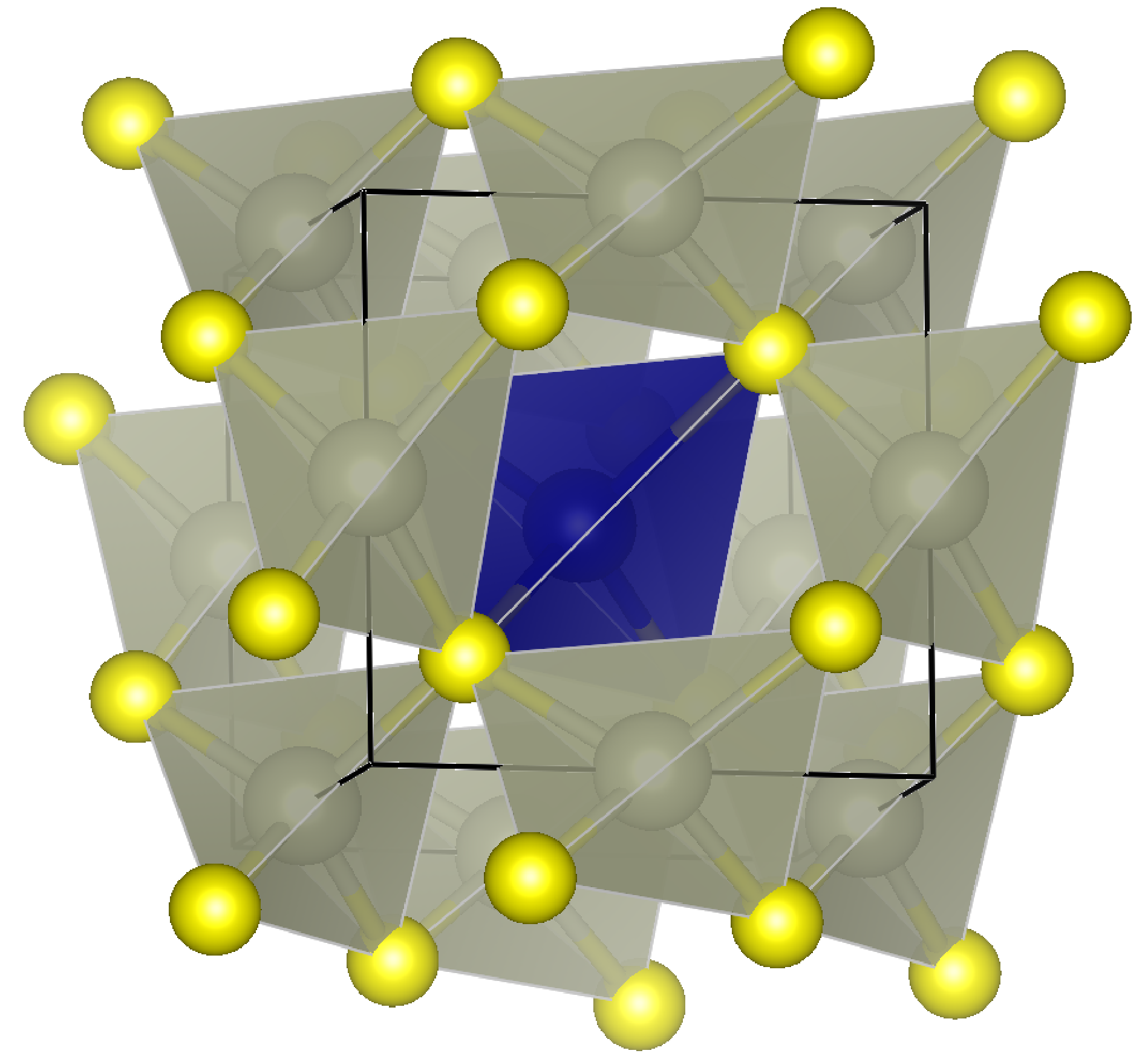}
		\caption{}
		\vspace{3pt}
	\end{subfigure}
	\begin{subfigure}{.32\textwidth}
		\includegraphics[width=\textwidth]{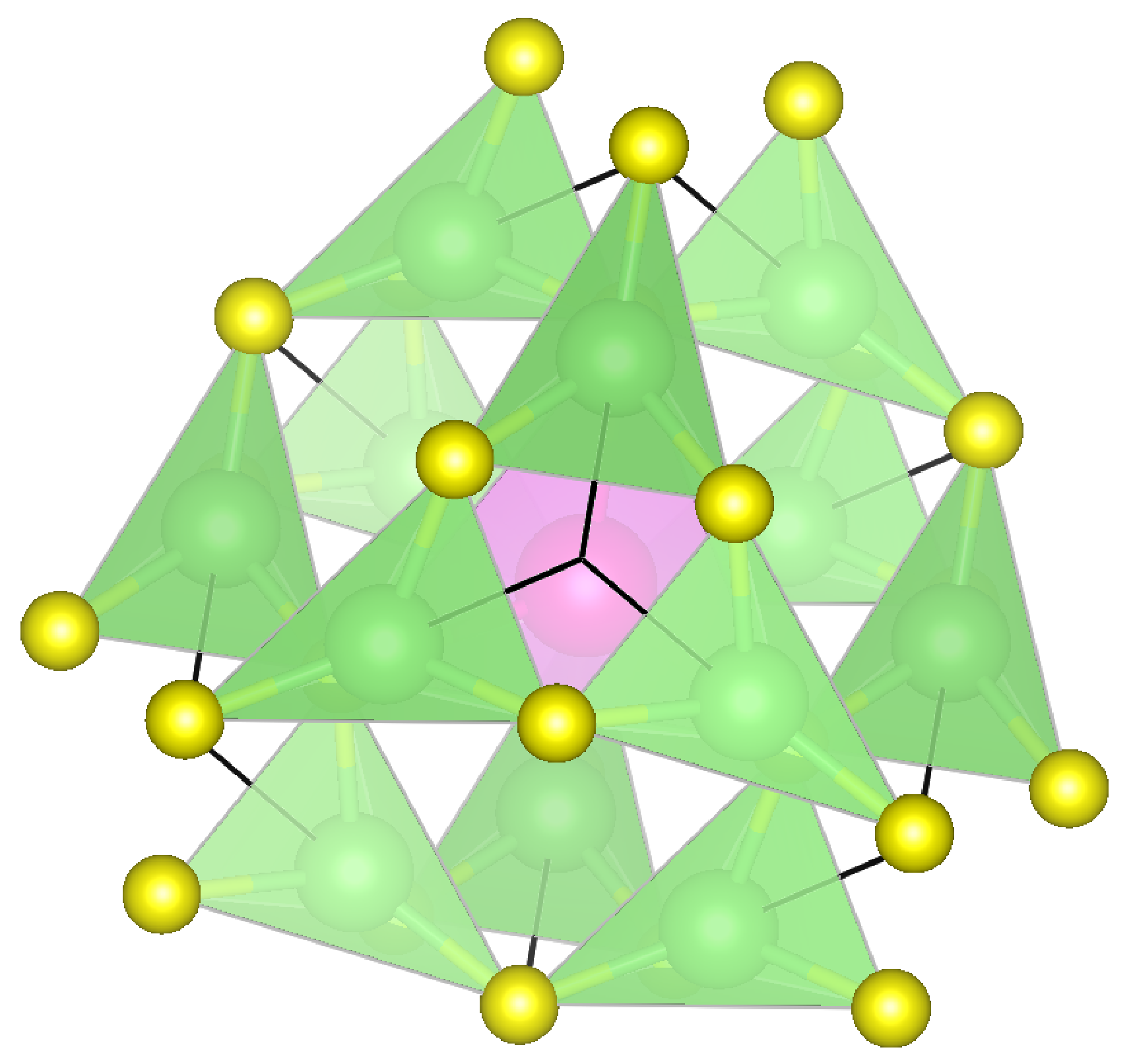}
		\caption{  }
		\vspace{3pt}
	\end{subfigure}
	\begin{subfigure}{.33\textwidth}
		\includegraphics[width=\textwidth]{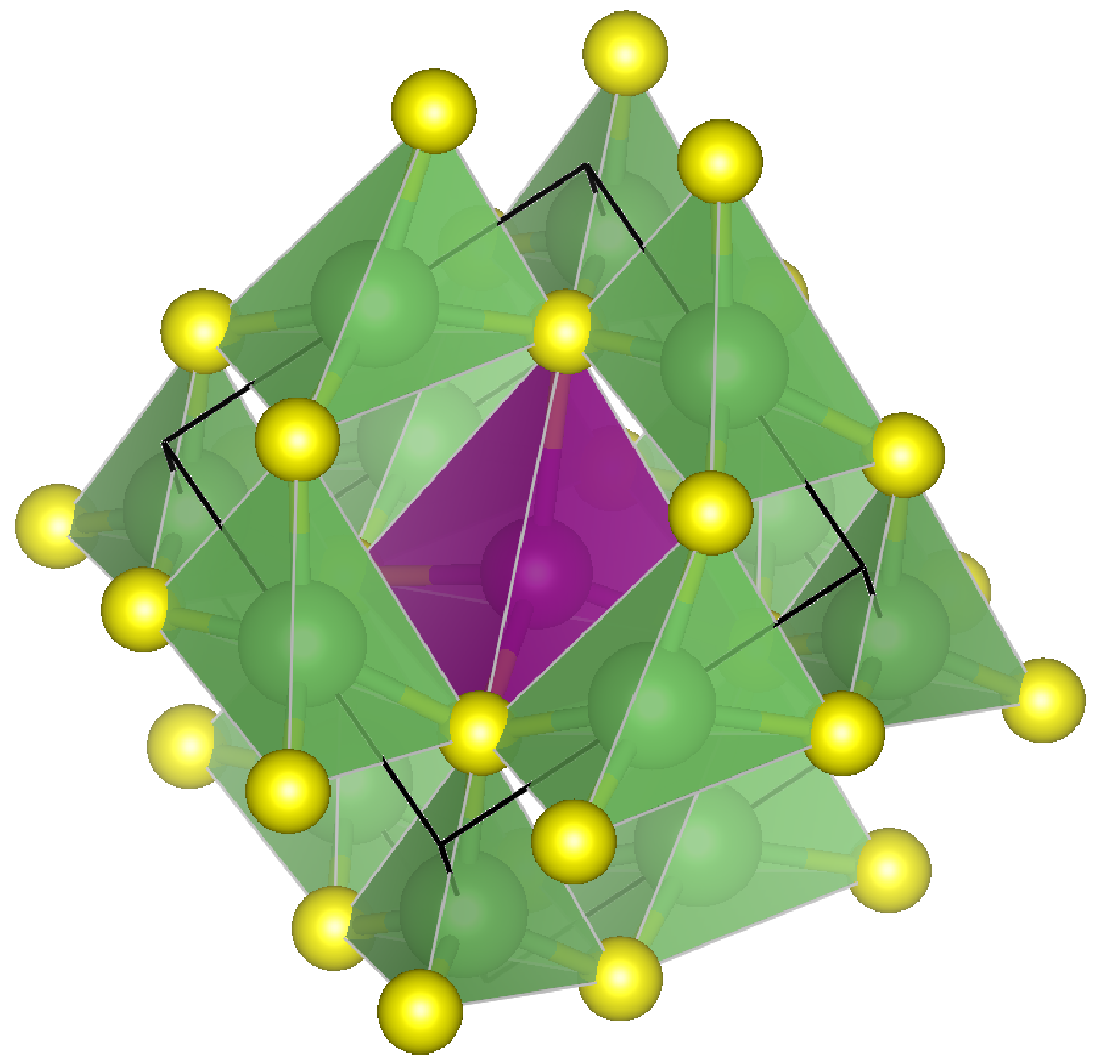}
		\caption{ }
		\vspace{3pt}
	\end{subfigure}
	\begin{subfigure}{.33\textwidth}
		\includegraphics[width=\textwidth]{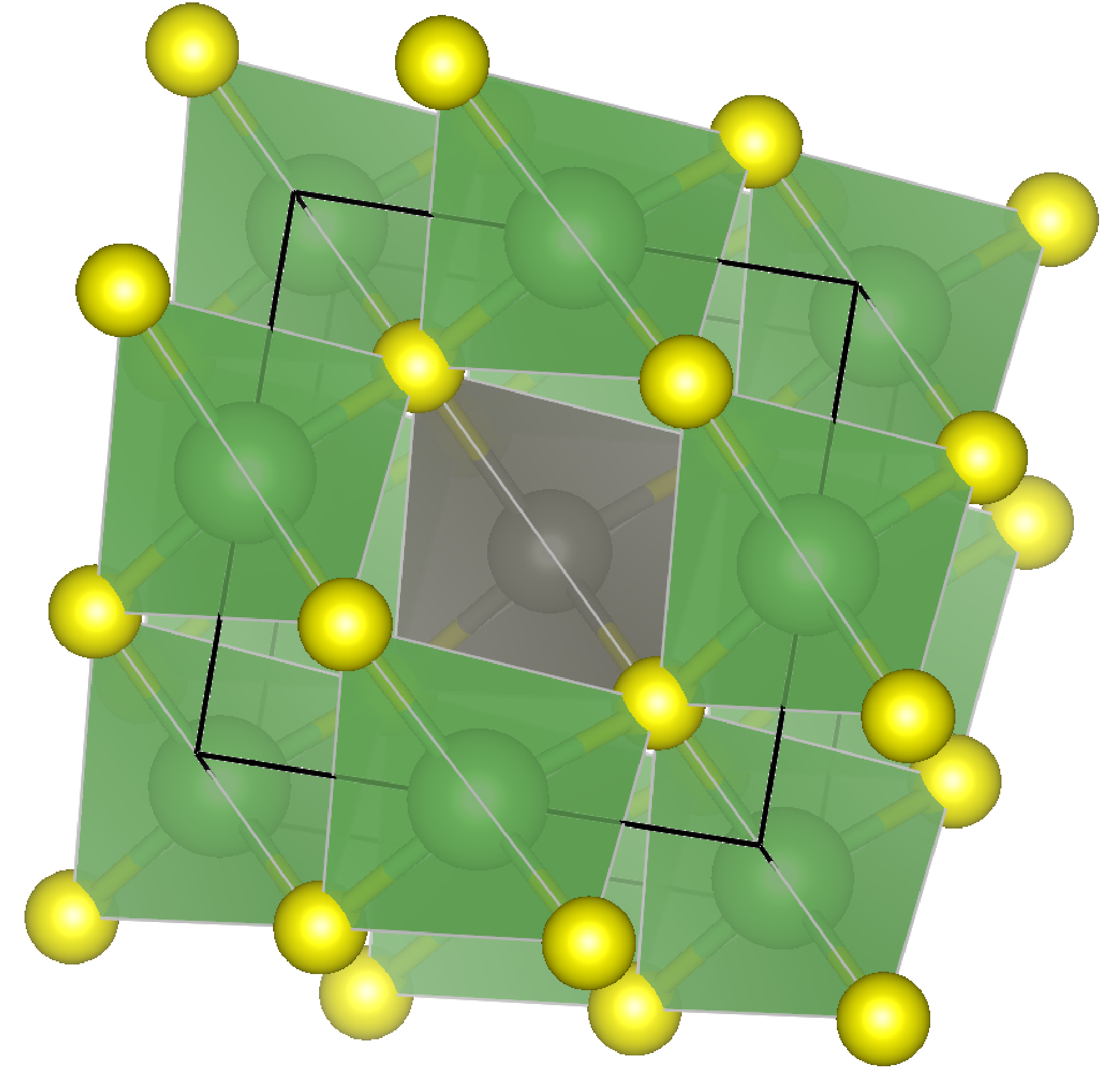}
		\caption{  }
		\vspace{3pt}
	\end{subfigure}
	
	\caption{Predicted crystal structures after DFT relaxation by AlphaCrystal. (a) Al\textsubscript{3}As\textsubscript{4}(Eform: -0.045 eV/atom). (b) CrCu\textsubscript{3}S\textsubscript{4} (Eform: -0.356 eV/atom (c) CrRh\textsubscript{3}S\textsubscript{4} (Eform: -0.287 eV/atom). (d) Li\textsubscript{3}LaS\textsubscript{4} (Eform: -0.991 eV/atom). (e) Li\textsubscript{3}MnS\textsubscript{4} (Eform: -0.822 eV/atom). (f) Li\textsubscript{3}ZnS\textsubscript{4} (Eform: -0.481 eV/atom)}
	\label{fig:predictedstructures}
\end{figure*}

For the above 100x10=1000 candidate structures, we use the contact map predictor to predict their contact maps and use the CMCrystal \cite{hu2020contact} algorithm to predict their crystal structures. For each of the 10 candidate structures of a given formula, we use graph neural network based model to predict their formation energy and pick the structure with minimum formation energy as its final structure. Out of the 100 predicted structures, we pick the top 7 with the lowest predicted formation energy to do DFT relaxation and phonon calculation to further determine their stability. The structures of these calculations are shown in Figure\ref{fig:predictedstructures}. Out of the 7 candidate structures, one structure is dynamically stable as shown in Fig.~\ref{fig:predictedstructures}, where there are no imaginary phonon frequencies.

\begin{table*}[bth]
\centering
\caption{Structural information and formation energy of six predicted new structures}
\label{tab:formationenergy}
\begin{tabular}{|l|r|r|r|r|r|r|r|l|}
\hline
\textbf{Material}              & \multicolumn{1}{l|}{space group} & \multicolumn{1}{l|}{a} & \multicolumn{1}{l|}{b} & \multicolumn{1}{l|}{c} & \multicolumn{1}{l|}{alpha} & \multicolumn{1}{l|}{beta} & \multicolumn{1}{l|}{gamma} & \textbf{Eform (ev/atom)} \\ \hline
{ Al3As4}  & 215                              & 5.3666                 & 5.3666                 & 5.3666                 & 90                         & 90                        & 90                         & -0.045         \\ \hline
{ CrCu3S4} & 215                              & 5.459                  & 5.459                  & 5.459                  & 90                         & 90                        & 90                         & -0.356         \\ \hline
{ CrRh3S4} & 164                              & 3.5322                 & 3.5322                 & 11.3559                & 90                         & 90                        & 120                        & -0.287         \\ \hline
{ Ge3P4}   & 164                              & 3.9036                 & 3.9036                 & 14.2124                & 90                         & 90                        & 120                        & 0.057          \\ \hline
{ Li3LaS4} & 225                              & 6.685                  & 6.685                  & 6.685                  & 90                         & 90                        & 90                         & -0.991         \\ \hline
{ Li3MnS4} & 225                              & 8.3957                 & 8.3957                 & 8.3957                 & 90                         & 90                        & 90                         & -0.822         \\ \hline
{ Li3ZnS4} & 215                              & 5.7787                 & 5.7787                 & 5.7787                 & 90                         & 90                        & 90                         & -0.481         \\ \hline
\end{tabular}
\end{table*}

\begin{figure*}[htb]
	\centering
	\begin{subfigure}{.33\textwidth}
		\includegraphics[width=\textwidth]{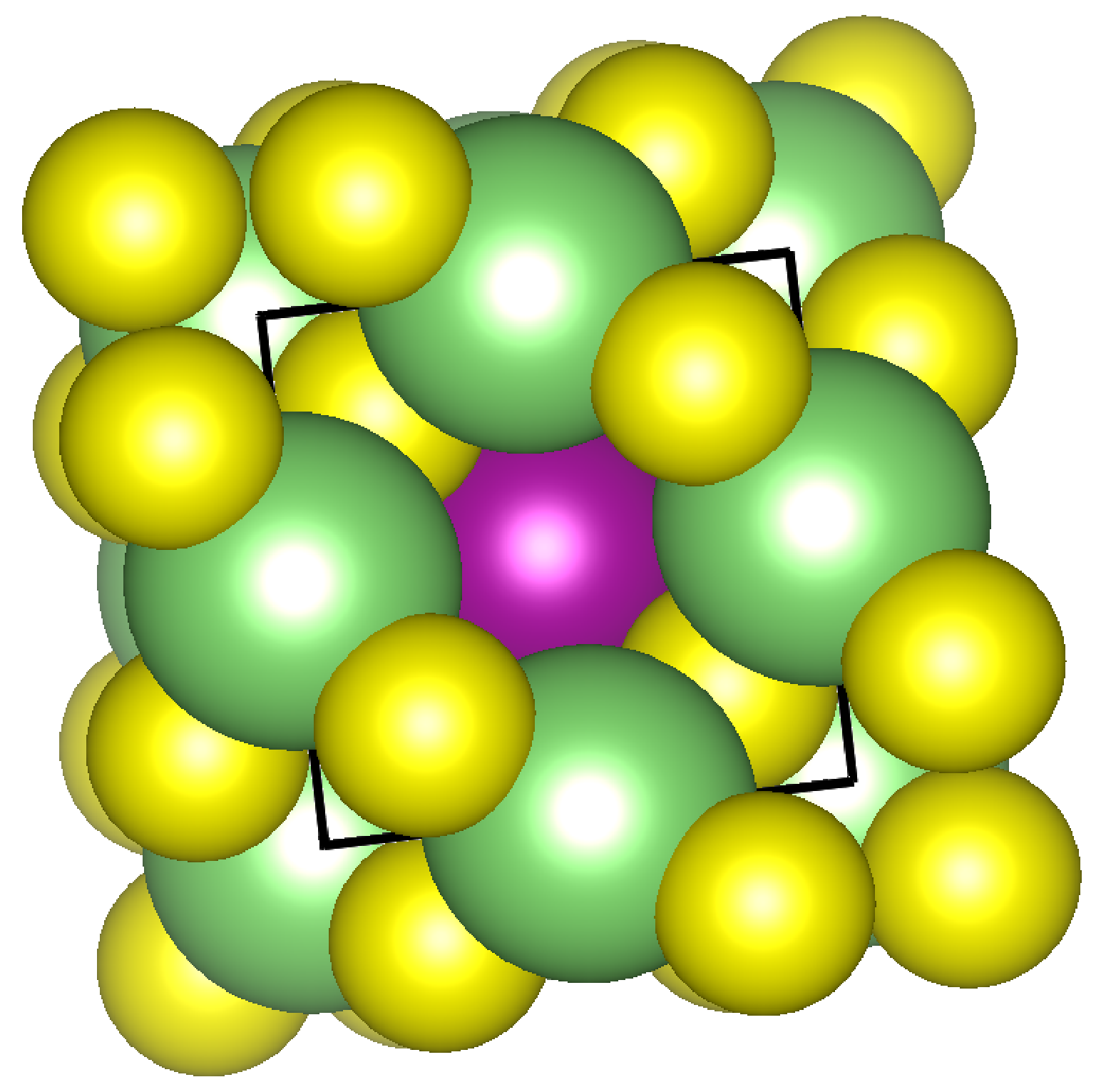}
		\caption{ }
		\vspace{3pt}
	\end{subfigure}
	\begin{subfigure}{.48\textwidth}
		\includegraphics[width=\textwidth]{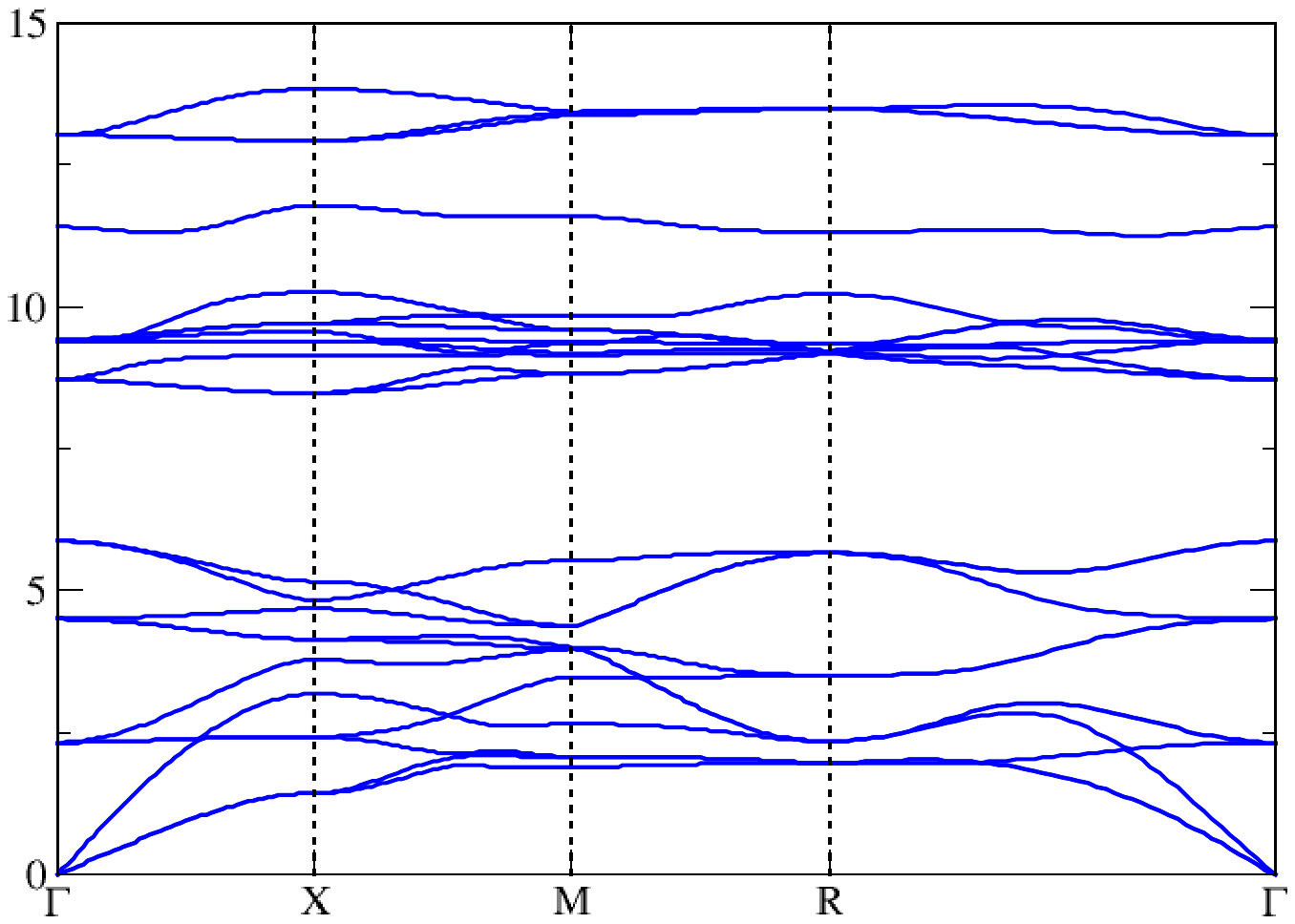}
		\caption{ }
		\vspace{3pt}
	\end{subfigure}
	\caption{Structure and phonon dispersion of Li\textsubscript{3}MnS\textsubscript{4} which is likely to be thermodynamically stable. (a) Structure (formation energy: -0.822 eV/atom). (b) Phonon dispersion}
	\label{fig:phonon}
\end{figure*}

\FloatBarrier

\section{Conclusion}
\label{sec:others}

We propose AlphaCrystal, a deep residual neural network approach for crystal structure prediction by first predicting the contact map of atom pairs for a given material composition, and then using it to predict its crystal structure using a genetic algorithm. Compared to the minimization of free energy during atomic configuration search in conventional ab initio CSP methods, our method takes advantage of the existing physical or geometric constraints (such as the symmetry of atom positions) of the existing crystal structures in the materials repositories. Our experiments show our AlphaCrystal algorithm is able to reconstruct the crystal structures for a large number of materials family of diverse space groups by optimizing the placement of the atoms using the contact map matching as the objective given only their space group and stoichiometry. We also applied y-scrambling to shuffle the structures of the compositions and found that the model trained with the shuffled dataset lost its contact map prediction power. Our predicted structures are so close to the target crystal structures so that they can be used to seed the costly free energy minimization based CSP algorithms for further structure refining. While we have demonstrated the feasibility of contact map based prediction of structures from formulas, we do recognize that the structure reconstruction from only the contact maps is not sufficient for successful structure prediction for many formulas. Adding distance constraints maybe next step. 
Overall, we believe our AlphaCrystal can be a new kind of deep knowledge guided approaches for large-scale prediction of crystal structures, which is very useful in high-throughput discovery of new materials using modern generative material design models \cite{fu2023material}.

\section{Availability of code and data}

The data that support the findings of this study are openly available in Materials Project database at \href{http:\\www.materialsproject.org}{\textcolor{blue}{http:\\www.materialsproject.org}}. The source code is available from \url{https://github.com/usccolumbia/AlphaCrystal}.

\section{Contribution}
Conceptualization, J.H.; methodology, J.H., Y.Z, W.Y., Q.L; software, Y.Z,W.Y., J.H, Q.L., and Y.S. ; validation, J.H., E.S., Y.Z., W.Y.;  investigation, J.H., Y.Z., W.Y., Q.L., Y.S., E.S., R.D.; resources, J.H.; data curation, J.H., Y.Z., and W.Y.; writing--original draft preparation, J.H.,  Y.Z., E.S.; writing--review and editing, J.H, Y.Z., E.S., R.D.; visualization, J.H., Y.Z., W.Y, E.S.; supervision, J.H.;  funding acquisition, J.H.

\begin{acknowledgement}

Research reported in this work was supported in part by NSF under grants 2110033, 1940099 and 1905775. The views, perspective, and content do not necessarily represent the official views of NSF. 

\end{acknowledgement}

\bibliography{references}

\providecommand{\latin}[1]{#1}
\makeatletter
\providecommand{\doi}
  {\begingroup\let\do\@makeother\dospecials
  \catcode`\{=1 \catcode`\}=2 \doi@aux}
\providecommand{\doi@aux}[1]{\endgroup\texttt{#1}}
\makeatother
\providecommand*\mcitethebibliography{\thebibliography}
\csname @ifundefined\endcsname{endmcitethebibliography}
  {\let\endmcitethebibliography\endthebibliography}{}
\begin{mcitethebibliography}{45}
\providecommand*\natexlab[1]{#1}
\providecommand*\mciteSetBstSublistMode[1]{}
\providecommand*\mciteSetBstMaxWidthForm[2]{}
\providecommand*\mciteBstWouldAddEndPuncttrue
  {\def\EndOfBibitem{\unskip.}}
\providecommand*\mciteBstWouldAddEndPunctfalse
  {\let\EndOfBibitem\relax}
\providecommand*\mciteSetBstMidEndSepPunct[3]{}
\providecommand*\mciteSetBstSublistLabelBeginEnd[3]{}
\providecommand*\EndOfBibitem{}
\mciteSetBstSublistMode{f}
\mciteSetBstMaxWidthForm{subitem}{(\alph{mcitesubitemcount})}
\mciteSetBstSublistLabelBeginEnd
  {\mcitemaxwidthsubitemform\space}
  {\relax}
  {\relax}

\bibitem[Dan \latin{et~al.}(2020)Dan, Zhao, Li, Li, Hu, and
  Hu]{dan2020generative}
Dan,~Y.; Zhao,~Y.; Li,~X.; Li,~S.; Hu,~M.; Hu,~J. Generative adversarial
  networks (GAN) based efficient sampling of chemical composition space for
  inverse design of inorganic materials. \emph{npj Computational Materials}
  \textbf{2020}, \emph{6}, 1--7\relax
\mciteBstWouldAddEndPuncttrue
\mciteSetBstMidEndSepPunct{\mcitedefaultmidpunct}
{\mcitedefaultendpunct}{\mcitedefaultseppunct}\relax
\EndOfBibitem
\bibitem[Maddox(1988)]{maddox1988crystals}
Maddox,~J. Crystals from first principles. \emph{Nature} \textbf{1988},
  \emph{335}, 201--201\relax
\mciteBstWouldAddEndPuncttrue
\mciteSetBstMidEndSepPunct{\mcitedefaultmidpunct}
{\mcitedefaultendpunct}{\mcitedefaultseppunct}\relax
\EndOfBibitem
\bibitem[Jang \latin{et~al.}(2020)Jang, Gu, Noh, Kim, and
  Jung]{jang2020structure}
Jang,~J.; Gu,~G.~H.; Noh,~J.; Kim,~J.; Jung,~Y. Structure-Based
  Synthesizability Prediction of Crystals Using Partially Supervised Learning.
  \emph{Journal of the American Chemical Society} \textbf{2020}, \relax
\mciteBstWouldAddEndPunctfalse
\mciteSetBstMidEndSepPunct{\mcitedefaultmidpunct}
{}{\mcitedefaultseppunct}\relax
\EndOfBibitem
\bibitem[Frey \latin{et~al.}(2019)Frey, Wang, Vega~Bellido, Anasori, Gogotsi,
  and Shenoy]{frey2019prediction}
Frey,~N.~C.; Wang,~J.; Vega~Bellido,~G.~I.; Anasori,~B.; Gogotsi,~Y.;
  Shenoy,~V.~B. Prediction of synthesis of 2D metal carbides and nitrides
  (MXenes) and their precursors with positive and unlabeled machine learning.
  \emph{ACS nano} \textbf{2019}, \emph{13}, 3031--3041\relax
\mciteBstWouldAddEndPuncttrue
\mciteSetBstMidEndSepPunct{\mcitedefaultmidpunct}
{\mcitedefaultendpunct}{\mcitedefaultseppunct}\relax
\EndOfBibitem
\bibitem[Aykol \latin{et~al.}(2019)Aykol, Hegde, Hung, Suram, Herring,
  Wolverton, and Hummelsh{\o}j]{aykol2019network}
Aykol,~M.; Hegde,~V.~I.; Hung,~L.; Suram,~S.; Herring,~P.; Wolverton,~C.;
  Hummelsh{\o}j,~J.~S. Network analysis of synthesizable materials discovery.
  \emph{Nature communications} \textbf{2019}, \emph{10}, 1--7\relax
\mciteBstWouldAddEndPuncttrue
\mciteSetBstMidEndSepPunct{\mcitedefaultmidpunct}
{\mcitedefaultendpunct}{\mcitedefaultseppunct}\relax
\EndOfBibitem
\bibitem[Xie and Grossman(2018)Xie, and Grossman]{xie2018crystal}
Xie,~T.; Grossman,~J.~C. Crystal graph convolutional neural networks for an
  accurate and interpretable prediction of material properties. \emph{Physical
  review letters} \textbf{2018}, \emph{120}, 145301\relax
\mciteBstWouldAddEndPuncttrue
\mciteSetBstMidEndSepPunct{\mcitedefaultmidpunct}
{\mcitedefaultendpunct}{\mcitedefaultseppunct}\relax
\EndOfBibitem
\bibitem[Tsuneyuki \latin{et~al.}(1988)Tsuneyuki, Tsukada, Aoki, and
  Matsui]{tsuneyuki1988first}
Tsuneyuki,~S.; Tsukada,~M.; Aoki,~H.; Matsui,~Y. First-principles interatomic
  potential of silica applied to molecular dynamics. \emph{Physical Review
  Letters} \textbf{1988}, \emph{61}, 869\relax
\mciteBstWouldAddEndPuncttrue
\mciteSetBstMidEndSepPunct{\mcitedefaultmidpunct}
{\mcitedefaultendpunct}{\mcitedefaultseppunct}\relax
\EndOfBibitem
\bibitem[Bush \latin{et~al.}(1995)Bush, Catlow, and
  Battle]{bush1995evolutionary}
Bush,~T.; Catlow,~C. R.~A.; Battle,~P. Evolutionary programming techniques for
  predicting inorganic crystal structures. \emph{Journal of Materials
  Chemistry} \textbf{1995}, \emph{5}, 1269--1272\relax
\mciteBstWouldAddEndPuncttrue
\mciteSetBstMidEndSepPunct{\mcitedefaultmidpunct}
{\mcitedefaultendpunct}{\mcitedefaultseppunct}\relax
\EndOfBibitem
\bibitem[Glass \latin{et~al.}(2006)Glass, Oganov, and Hansen]{glass2006uspex}
Glass,~C.~W.; Oganov,~A.~R.; Hansen,~N. USPEX—Evolutionary crystal structure
  prediction. \emph{Computer physics communications} \textbf{2006}, \emph{175},
  713--720\relax
\mciteBstWouldAddEndPuncttrue
\mciteSetBstMidEndSepPunct{\mcitedefaultmidpunct}
{\mcitedefaultendpunct}{\mcitedefaultseppunct}\relax
\EndOfBibitem
\bibitem[Woodley and Catlow(2008)Woodley, and Catlow]{woodley2008crystal}
Woodley,~S.~M.; Catlow,~R. Crystal structure prediction from first principles.
  \emph{Nature materials} \textbf{2008}, \emph{7}, 937--946\relax
\mciteBstWouldAddEndPuncttrue
\mciteSetBstMidEndSepPunct{\mcitedefaultmidpunct}
{\mcitedefaultendpunct}{\mcitedefaultseppunct}\relax
\EndOfBibitem
\bibitem[Hautier \latin{et~al.}(2011)Hautier, Fischer, Ehrlacher, Jain, and
  Ceder]{hautier2011data}
Hautier,~G.; Fischer,~C.; Ehrlacher,~V.; Jain,~A.; Ceder,~G. Data mined ionic
  substitutions for the discovery of new compounds. \emph{Inorganic chemistry}
  \textbf{2011}, \emph{50}, 656--663\relax
\mciteBstWouldAddEndPuncttrue
\mciteSetBstMidEndSepPunct{\mcitedefaultmidpunct}
{\mcitedefaultendpunct}{\mcitedefaultseppunct}\relax
\EndOfBibitem
\bibitem[Wang \latin{et~al.}(2012)Wang, Lv, Zhu, and Ma]{wang2012calypso}
Wang,~Y.; Lv,~J.; Zhu,~L.; Ma,~Y. CALYPSO: A method for crystal structure
  prediction. \emph{Computer Physics Communications} \textbf{2012}, \emph{183},
  2063--2070\relax
\mciteBstWouldAddEndPuncttrue
\mciteSetBstMidEndSepPunct{\mcitedefaultmidpunct}
{\mcitedefaultendpunct}{\mcitedefaultseppunct}\relax
\EndOfBibitem
\bibitem[Curtis \latin{et~al.}(2018)Curtis, Li, Rose, Vazquez-Mayagoitia,
  Bhattacharya, Ghiringhelli, and Marom]{curtis2018gator}
Curtis,~F.; Li,~X.; Rose,~T.; Vazquez-Mayagoitia,~A.; Bhattacharya,~S.;
  Ghiringhelli,~L.~M.; Marom,~N. GAtor: a first-principles genetic algorithm
  for molecular crystal structure prediction. \emph{Journal of chemical theory
  and computation} \textbf{2018}, \emph{14}, 2246--2264\relax
\mciteBstWouldAddEndPuncttrue
\mciteSetBstMidEndSepPunct{\mcitedefaultmidpunct}
{\mcitedefaultendpunct}{\mcitedefaultseppunct}\relax
\EndOfBibitem
\bibitem[Avery \latin{et~al.}(2019)Avery, Toher, Curtarolo, and
  Zurek]{avery2019xtalopt}
Avery,~P.; Toher,~C.; Curtarolo,~S.; Zurek,~E. XtalOpt Version r12: An
  open-source evolutionary algorithm for crystal structure prediction.
  \emph{Comput. Phys. Commun.} \textbf{2019}, \emph{237}, 274--275\relax
\mciteBstWouldAddEndPuncttrue
\mciteSetBstMidEndSepPunct{\mcitedefaultmidpunct}
{\mcitedefaultendpunct}{\mcitedefaultseppunct}\relax
\EndOfBibitem
\bibitem[Ryan \latin{et~al.}(2018)Ryan, Lengyel, and Shatruk]{ryan2018crystal}
Ryan,~K.; Lengyel,~J.; Shatruk,~M. Crystal structure prediction via deep
  learning. \emph{Journal of the American Chemical Society} \textbf{2018},
  \emph{140}, 10158--10168\relax
\mciteBstWouldAddEndPuncttrue
\mciteSetBstMidEndSepPunct{\mcitedefaultmidpunct}
{\mcitedefaultendpunct}{\mcitedefaultseppunct}\relax
\EndOfBibitem
\bibitem[Oganov \latin{et~al.}(2019)Oganov, Pickard, Zhu, and
  Needs]{oganov2019structure}
Oganov,~A.~R.; Pickard,~C.~J.; Zhu,~Q.; Needs,~R.~J. Structure prediction
  drives materials discovery. \emph{Nature Reviews Materials} \textbf{2019},
  \emph{4}, 331--348\relax
\mciteBstWouldAddEndPuncttrue
\mciteSetBstMidEndSepPunct{\mcitedefaultmidpunct}
{\mcitedefaultendpunct}{\mcitedefaultseppunct}\relax
\EndOfBibitem
\bibitem[Lyakhov \latin{et~al.}(2013)Lyakhov, Oganov, Stokes, and
  Zhu]{lyakhov2013new}
Lyakhov,~A.~O.; Oganov,~A.~R.; Stokes,~H.~T.; Zhu,~Q. New developments in
  evolutionary structure prediction algorithm USPEX. \emph{Computer Physics
  Communications} \textbf{2013}, \emph{184}, 1172--1182\relax
\mciteBstWouldAddEndPuncttrue
\mciteSetBstMidEndSepPunct{\mcitedefaultmidpunct}
{\mcitedefaultendpunct}{\mcitedefaultseppunct}\relax
\EndOfBibitem
\bibitem[Fischer \latin{et~al.}(2006)Fischer, Tibbetts, Morgan, and
  Ceder]{fischer2006predicting}
Fischer,~C.~C.; Tibbetts,~K.~J.; Morgan,~D.; Ceder,~G. Predicting crystal
  structure by merging data mining with quantum mechanics. \emph{Nature
  materials} \textbf{2006}, \emph{5}, 641--646\relax
\mciteBstWouldAddEndPuncttrue
\mciteSetBstMidEndSepPunct{\mcitedefaultmidpunct}
{\mcitedefaultendpunct}{\mcitedefaultseppunct}\relax
\EndOfBibitem
\bibitem[Shen \latin{et~al.}(2020)Shen, Horton, and Persson]{shen2020charge}
Shen,~J.-X.; Horton,~M.; Persson,~K.~A. A charge-density-based general cation
  insertion algorithm for generating new Li-ion cathode materials. \emph{npj
  Computational Materials} \textbf{2020}, \emph{6}, 1--7\relax
\mciteBstWouldAddEndPuncttrue
\mciteSetBstMidEndSepPunct{\mcitedefaultmidpunct}
{\mcitedefaultendpunct}{\mcitedefaultseppunct}\relax
\EndOfBibitem
\bibitem[He \latin{et~al.}(2020)He, Yao, Hegde, Naghavi, Shen, Bushick, and
  Wolverton]{he2020computational}
He,~J.; Yao,~Z.; Hegde,~V.~I.; Naghavi,~S.~S.; Shen,~J.; Bushick,~K.~M.;
  Wolverton,~C. Computational Discovery of Stable Heteroanionic
  Oxychalcogenides ABXO (A, B= Metals; X= S, Se, and Te) and Their Potential
  Applications. \emph{Chemistry of Materials} \textbf{2020}, \emph{32},
  8229--8242\relax
\mciteBstWouldAddEndPuncttrue
\mciteSetBstMidEndSepPunct{\mcitedefaultmidpunct}
{\mcitedefaultendpunct}{\mcitedefaultseppunct}\relax
\EndOfBibitem
\bibitem[Wang \latin{et~al.}(2020)Wang, Lv, Li, Wang, and Ma]{wang2020calypso}
Wang,~Y.; Lv,~J.; Li,~Q.; Wang,~H.; Ma,~Y. CALYPSO method for structure
  prediction and its applications to materials discovery. \emph{Handbook of
  Materials Modeling: Applications: Current and Emerging Materials}
  \textbf{2020}, 2729--2756\relax
\mciteBstWouldAddEndPuncttrue
\mciteSetBstMidEndSepPunct{\mcitedefaultmidpunct}
{\mcitedefaultendpunct}{\mcitedefaultseppunct}\relax
\EndOfBibitem
\bibitem[Pretti \latin{et~al.}(2020)Pretti, Shen, Mittal, and
  Mahynski]{pretti2020symmetry}
Pretti,~E.; Shen,~V.~K.; Mittal,~J.; Mahynski,~N.~A. Symmetry-Based Crystal
  Structure Enumeration in Two Dimensions. \emph{The Journal of Physical
  Chemistry A} \textbf{2020}, \emph{124}, 3276--3285\relax
\mciteBstWouldAddEndPuncttrue
\mciteSetBstMidEndSepPunct{\mcitedefaultmidpunct}
{\mcitedefaultendpunct}{\mcitedefaultseppunct}\relax
\EndOfBibitem
\bibitem[Podryabinkin \latin{et~al.}(2019)Podryabinkin, Tikhonov, Shapeev, and
  Oganov]{podryabinkin2019accelerating}
Podryabinkin,~E.~V.; Tikhonov,~E.~V.; Shapeev,~A.~V.; Oganov,~A.~R.
  Accelerating crystal structure prediction by machine-learning interatomic
  potentials with active learning. \emph{Physical Review B} \textbf{2019},
  \emph{99}, 064114\relax
\mciteBstWouldAddEndPuncttrue
\mciteSetBstMidEndSepPunct{\mcitedefaultmidpunct}
{\mcitedefaultendpunct}{\mcitedefaultseppunct}\relax
\EndOfBibitem
\bibitem[Zhang \latin{et~al.}(2017)Zhang, Wang, Lv, and Ma]{zhang2017materials}
Zhang,~L.; Wang,~Y.; Lv,~J.; Ma,~Y. Materials discovery at high pressures.
  \emph{Nature Reviews Materials} \textbf{2017}, \emph{2}, 1--16\relax
\mciteBstWouldAddEndPuncttrue
\mciteSetBstMidEndSepPunct{\mcitedefaultmidpunct}
{\mcitedefaultendpunct}{\mcitedefaultseppunct}\relax
\EndOfBibitem
\bibitem[Di~Lena \latin{et~al.}(2012)Di~Lena, Nagata, and Baldi]{di2012deep}
Di~Lena,~P.; Nagata,~K.; Baldi,~P. Deep architectures for protein contact map
  prediction. \emph{Bioinformatics} \textbf{2012}, \emph{28}, 2449--2457\relax
\mciteBstWouldAddEndPuncttrue
\mciteSetBstMidEndSepPunct{\mcitedefaultmidpunct}
{\mcitedefaultendpunct}{\mcitedefaultseppunct}\relax
\EndOfBibitem
\bibitem[Senior \latin{et~al.}(2020)Senior, Evans, Jumper, Kirkpatrick, Sifre,
  Green, Qin, {\v{Z}}{\'\i}dek, Nelson, Bridgland, \latin{et~al.}
  others]{senior2020improved}
others,, \latin{et~al.}  Improved protein structure prediction using potentials
  from deep learning. \emph{Nature} \textbf{2020}, \emph{577}, 706--710\relax
\mciteBstWouldAddEndPuncttrue
\mciteSetBstMidEndSepPunct{\mcitedefaultmidpunct}
{\mcitedefaultendpunct}{\mcitedefaultseppunct}\relax
\EndOfBibitem
\bibitem[Zheng \latin{et~al.}(2019)Zheng, Li, Zhang, Pearce, Mortuza, and
  Zhang]{zheng2019deep}
Zheng,~W.; Li,~Y.; Zhang,~C.; Pearce,~R.; Mortuza,~S.; Zhang,~Y. Deep-learning
  contact-map guided protein structure prediction in CASP13. \emph{Proteins:
  Structure, Function, and Bioinformatics} \textbf{2019}, \emph{87},
  1149--1164\relax
\mciteBstWouldAddEndPuncttrue
\mciteSetBstMidEndSepPunct{\mcitedefaultmidpunct}
{\mcitedefaultendpunct}{\mcitedefaultseppunct}\relax
\EndOfBibitem
\bibitem[He \latin{et~al.}(2016)He, Zhang, Ren, and Sun]{he2016identity}
He,~K.; Zhang,~X.; Ren,~S.; Sun,~J. Identity mappings in deep residual
  networks. European conference on computer vision. 2016; pp 630--645\relax
\mciteBstWouldAddEndPuncttrue
\mciteSetBstMidEndSepPunct{\mcitedefaultmidpunct}
{\mcitedefaultendpunct}{\mcitedefaultseppunct}\relax
\EndOfBibitem
\bibitem[Zhu \latin{et~al.}(2017)Zhu, Wu, Wu, Xu, Xu, Zhao, Wang, and
  Ho]{zhu2017efficient}
Zhu,~Z.; Wu,~P.; Wu,~S.; Xu,~L.; Xu,~Y.; Zhao,~X.; Wang,~C.-Z.; Ho,~K.-M. An
  Efficient Scheme for Crystal Structure Prediction Based on Structural Motifs.
  \emph{The Journal of Physical Chemistry C} \textbf{2017}, \emph{121},
  11891--11896\relax
\mciteBstWouldAddEndPuncttrue
\mciteSetBstMidEndSepPunct{\mcitedefaultmidpunct}
{\mcitedefaultendpunct}{\mcitedefaultseppunct}\relax
\EndOfBibitem
\bibitem[Hu \latin{et~al.}(2020)Hu, Yang, Dong, Li, Li, and Li]{hu2020contact}
Hu,~J.; Yang,~W.; Dong,~R.; Li,~Y.; Li,~X.; Li,~S. Contact Map based Crystal
  Structure Prediction using Global Optimization. \emph{arXiv preprint
  arXiv:2008.07016} \textbf{2020}, \relax
\mciteBstWouldAddEndPunctfalse
\mciteSetBstMidEndSepPunct{\mcitedefaultmidpunct}
{}{\mcitedefaultseppunct}\relax
\EndOfBibitem
\bibitem[Liang \latin{et~al.}(2020)Liang, Stanev, Kusne, and
  Takeuchi]{liang2020cryspnet}
Liang,~H.; Stanev,~V.; Kusne,~A.~G.; Takeuchi,~I. CRYSPNet: Crystal Structure
  Predictions via Neural Network. \emph{arXiv preprint arXiv:2003.14328}
  \textbf{2020}, \relax
\mciteBstWouldAddEndPunctfalse
\mciteSetBstMidEndSepPunct{\mcitedefaultmidpunct}
{}{\mcitedefaultseppunct}\relax
\EndOfBibitem
\bibitem[Li \latin{et~al.}(2020)Li, Yang, Dong, and Hu]{li2020mlatticeabc}
Li,~Y.; Yang,~W.; Dong,~R.; Hu,~J. MLatticeABC: Generic Lattice Constant
  Prediction of Crystal Materials using Machine Learning. \emph{arXiv preprint
  arXiv:2010.16099} \textbf{2020}, \relax
\mciteBstWouldAddEndPunctfalse
\mciteSetBstMidEndSepPunct{\mcitedefaultmidpunct}
{}{\mcitedefaultseppunct}\relax
\EndOfBibitem
\bibitem[Ioffe and Szegedy(2015)Ioffe, and Szegedy]{ioffe2015batch}
Ioffe,~S.; Szegedy,~C. Batch normalization: Accelerating deep network training
  by reducing internal covariate shift. \emph{arXiv preprint arXiv:1502.03167}
  \textbf{2015}, \relax
\mciteBstWouldAddEndPunctfalse
\mciteSetBstMidEndSepPunct{\mcitedefaultmidpunct}
{}{\mcitedefaultseppunct}\relax
\EndOfBibitem
\bibitem[Krizhevsky \latin{et~al.}(2017)Krizhevsky, Sutskever, and
  Hinton]{krizhevsky2017imagenet}
Krizhevsky,~A.; Sutskever,~I.; Hinton,~G.~E. Imagenet classification with deep
  convolutional neural networks. \emph{Communications of the ACM}
  \textbf{2017}, \emph{60}, 84--90\relax
\mciteBstWouldAddEndPuncttrue
\mciteSetBstMidEndSepPunct{\mcitedefaultmidpunct}
{\mcitedefaultendpunct}{\mcitedefaultseppunct}\relax
\EndOfBibitem
\bibitem[Kresse and Hafner(1993)Kresse, and Hafner]{Vasp1}
Kresse,~G.; Hafner,~J. ab initio. \emph{Phys. Rev. B} \textbf{1993}, \emph{47},
  558--561\relax
\mciteBstWouldAddEndPuncttrue
\mciteSetBstMidEndSepPunct{\mcitedefaultmidpunct}
{\mcitedefaultendpunct}{\mcitedefaultseppunct}\relax
\EndOfBibitem
\bibitem[Kresse and Hafner(1994)Kresse, and Hafner]{Vasp2}
Kresse,~G.; Hafner,~J. ab initio. \emph{Phys. Rev. B} \textbf{1994}, \emph{49},
  14251--14269\relax
\mciteBstWouldAddEndPuncttrue
\mciteSetBstMidEndSepPunct{\mcitedefaultmidpunct}
{\mcitedefaultendpunct}{\mcitedefaultseppunct}\relax
\EndOfBibitem
\bibitem[G.~Kresse(1996)]{Vasp3}
G.~Kresse,~J.~F. Efficiency of ab initio Total Energy Calculations for Metals
  and Semiconductors Using a Plane-Wave Basis Set. \emph{Comput. Mater. Sci.}
  \textbf{1996}, \emph{6}, 15--50\relax
\mciteBstWouldAddEndPuncttrue
\mciteSetBstMidEndSepPunct{\mcitedefaultmidpunct}
{\mcitedefaultendpunct}{\mcitedefaultseppunct}\relax
\EndOfBibitem
\bibitem[Kresse and Furthm\"uller(1996)Kresse, and Furthm\"uller]{Vasp4}
Kresse,~G.; Furthm\"uller,~J. Efficient Iterative Schemes for ab initio
  Total-Energy Calculations Using a Plane-Wave Basis Set. \emph{Phys. Rev. B}
  \textbf{1996}, \emph{54}, 11169--11186\relax
\mciteBstWouldAddEndPuncttrue
\mciteSetBstMidEndSepPunct{\mcitedefaultmidpunct}
{\mcitedefaultendpunct}{\mcitedefaultseppunct}\relax
\EndOfBibitem
\bibitem[Bl\"ochl(1994)]{PAW1}
Bl\"ochl,~P.~E. Projector Augmented-Wave Method. \emph{Phys. Rev. B}
  \textbf{1994}, \emph{50}, 17953--17979\relax
\mciteBstWouldAddEndPuncttrue
\mciteSetBstMidEndSepPunct{\mcitedefaultmidpunct}
{\mcitedefaultendpunct}{\mcitedefaultseppunct}\relax
\EndOfBibitem
\bibitem[Kresse and Joubert(1999)Kresse, and Joubert]{PAW2}
Kresse,~G.; Joubert,~D. From Ultrasoft Pseudopotentials to the Projector
  Augmented-Wave Method. \emph{Phys. Rev. B} \textbf{1999}, \emph{59},
  1758--1775\relax
\mciteBstWouldAddEndPuncttrue
\mciteSetBstMidEndSepPunct{\mcitedefaultmidpunct}
{\mcitedefaultendpunct}{\mcitedefaultseppunct}\relax
\EndOfBibitem
\bibitem[Perdew \latin{et~al.}(1996)Perdew, Burke, and Ernzerhof]{GGA1}
Perdew,~J.~P.; Burke,~K.; Ernzerhof,~M. Generalized Gradient Approximation Made
  Simple. \emph{Phys. Rev. Lett.} \textbf{1996}, \emph{77}, 3865--3868\relax
\mciteBstWouldAddEndPuncttrue
\mciteSetBstMidEndSepPunct{\mcitedefaultmidpunct}
{\mcitedefaultendpunct}{\mcitedefaultseppunct}\relax
\EndOfBibitem
\bibitem[Perdew \latin{et~al.}(1997)Perdew, Burke, and Ernzerhof]{GGA2}
Perdew,~J.~P.; Burke,~K.; Ernzerhof,~M. Generalized Gradient Approximation Made
  Simple [Phys. Rev. Lett. 77, 3865 (1996)]. \emph{Phys. Rev. Lett.}
  \textbf{1997}, \emph{78}, 1396--1396\relax
\mciteBstWouldAddEndPuncttrue
\mciteSetBstMidEndSepPunct{\mcitedefaultmidpunct}
{\mcitedefaultendpunct}{\mcitedefaultseppunct}\relax
\EndOfBibitem
\bibitem[Kingma and Ba(2014)Kingma, and Ba]{kingma2014adam}
Kingma,~D.~P.; Ba,~J. Adam: A method for stochastic optimization. \emph{arXiv
  preprint arXiv:1412.6980} \textbf{2014}, \relax
\mciteBstWouldAddEndPunctfalse
\mciteSetBstMidEndSepPunct{\mcitedefaultmidpunct}
{}{\mcitedefaultseppunct}\relax
\EndOfBibitem
\bibitem[Fu \latin{et~al.}(2023)Fu, Wei, Song, Li, Xin, Omee, Dong,
  Siriwardane, and Hu]{fu2023material}
Fu,~N.; Wei,~L.; Song,~Y.; Li,~Q.; Xin,~R.; Omee,~S.~S.; Dong,~R.;
  Siriwardane,~E. M.~D.; Hu,~J. Material transformers: deep learning language
  models for generative materials design. \emph{Machine Learning: Science and
  Technology} \textbf{2023}, \emph{4}, 015001\relax
\mciteBstWouldAddEndPuncttrue
\mciteSetBstMidEndSepPunct{\mcitedefaultmidpunct}
{\mcitedefaultendpunct}{\mcitedefaultseppunct}\relax
\EndOfBibitem
\end{mcitethebibliography}

\end{document}